\documentclass[
 reprint,
 amsmath,amssymb,
 aps,
]{revtex4-2}

\usepackage{braket}
\usepackage{booktabs}
\usepackage{graphicx}
\usepackage{dcolumn}
\usepackage{bm}
\usepackage{xcolor}
\usepackage{svg}
\usepackage[caption=false]{subfig}
\usepackage{comment}
\usepackage{orcidlink}

\usepackage{glossaries}
\newacronym{bvk}{BvK}{Born-von-Karman} 
\newacronym{hf}{HF}{Hartree--Fock} 
\newacronym{tdl}{TDL}{thermodynamic limit} 
\usepackage[]{hyperref}

\begin{document}

\preprint{APS/123-QED}

\title{
  Finite-size Effects in periodic EOM-CCSD for Ionization Energies and Electron Affinities: 
  Convergence Rate and Extrapolation to the Thermodynamic Limit}

\def\vienna{%
  \affiliation{%
    Institute for Theoretical Physics, TU Wien,\\
    Wiedner Hauptstra{\ss}e 8--10/136, 1040 Vienna, Austria 
  }%
}

\def\nomadaff{%
  \affiliation{%
    The NOMAD Laboratory at the FHI of the Max-Planck-Gesellschaft%
  }
}

\author{Evgeny Moerman\,\orcidlink{0000-0002-9725-612X}}
\email{moerman@fhi-berlin.mpg.de}
\nomadaff
\author{Alejandro Gallo\,\orcidlink{0000-0002-3744-4161}}
\vienna
\author{Andreas Irmler\,\orcidlink{0000-0003-0525-7772}}
\vienna
\author{Tobias Sch\"{a}fer\,\orcidlink{0000-0003-0716-6516}}
\vienna
\author{Felix Hummel\,\orcidlink{0000-0003-1941-3385}}
\vienna
\author{Andreas Gr\"{u}neis\,\orcidlink{0000-0002-4984-7785}}
\vienna
\author{Matthias Scheffler\,\orcidlink{0000-0002-1280-9873}}
\nomadaff

\date{\today}

\begin{abstract}
We investigate the convergence of quasi-particle energies
for periodic systems to the thermodynamic limit using increasingly 
large simulation cells corresponding to increasingly dense integration meshes in reciprocal space.
The quasi-particle energies are computed at the level of
equation-of-motion coupled-cluster theory for ionization
(IP-EOM-CC) and electron attachment processes (EA-EOM-CC).
By introducing an electronic correlation structure factor, the
expected asymptotic convergence rates for systems with different dimensionality
are formally derived.
We rigorously test these derivations through numerical simulations for 
\textit{trans}-Polyacetylene using IP/EA-EOM-CCSD and the $G_0W_0$@HF approximation, 
which confirm the predicted convergence behavior.
Our findings provide a solid foundation for efficient schemes to correct 
finite-size errors in IP/EA-EOM-CCSD calculations.
\end{abstract}

\maketitle

\section{\label{sec:intro}Introduction}

For most theoretical materials science studies, density functional
theory (DFT) is employed due to its favorable balance between
computational scaling and moderate accuracy. However, for many materials
properties the accurate inclusion of electronic exchange and correlation
effects is critical to achieve a qualitatively correct description for
scientifically and technologically important properties. In particular,
for the theoretical description
of electronic band gaps and band structures, most Kohn-Sham density functional
approximations (KS-DFAs) in use today are known to generally severely
underestimate
band gaps, oftentimes referred to as the \textit{band gap problem}~\cite{perdew2017understanding}.
Relaxing the condition of a multiplicative exchange-correlation potential of KS-DFT to allow
for a more flexible integral operator as it is the case in generalized KS-DFT resolves many of 
the issues associated with the band gap problem~\cite{perdew2017understanding}, 
generally reducing the band gap error and rectifying the missing derivative discontinuity.
Apart from hybrid functionals, which are the most widely used representatives 
of generalized KS-DFAs, the electronic structure method of choice for the
calculation of quasi-particle energies has become the
$GW$-approximation~\cite{hedin1965new,golze2019gw}, which takes the DFA
one-electron wave functions as a starting point but explicitly
accounts for electronic exchange and
correlation effects using a perturbation theory approach.
The $GW$-approximation yields significantly improved band gaps
compared to the most widely used approximate density functionals.
However, the most commonly used method based on the $GW$ approximation, the
$G_0W_0$ method,
is known to have its limitations as well, the most glaring one being
the dependency of the band gap result on the underlying DFA, the so-called
\textit{starting point dependence}~\cite{golze2019gw}.
Other higher-order corrections to the $GW$ approximation require
the inclusion of so-called vertex corrections in the self-energy and
the screened interaction $W$~\cite{shishkin2007accurate,Gruneis2014}.
Although certain improvements can be achieved
using vertex corrections, it remains challenging to systematically improve
upon the $GW$ approximation, which already achieves an excellent trade-off
between computational cost and accuracy~\cite{lewis2019vertex, kutepov2022full, romaniello2012beyond}.

A systematically improvable method is the equation-of-motion
coupled-cluster~\cite{stanton1993equation}(EOM-CC) framework.
EOM-CC theory, being the extension of ground-state CC theory to excited states, allows to
theoretically describe systems upon removal (IP-EOM-CC), addition
(EA-EOM-CC) or vertical excitation (EE-EOM-CC) of an electron. As the
electronic band gap is defined as the difference between the electron
affinity/attachment (EA) and the ionization potential (IP), it is possible
to obtain band gaps and entire band structures in the EOM-CC
framework~\cite{mcclain2017gaussian,gao2020electronic}.
The relation between EOM-CC and $GW$ band gaps was also investigated,
showing the differences and similiarites between the diagramatic contributions 
and the results of both approaches~\cite{lange2018relation,mcclain2016spectral, tolle2023exact}.
However, the major obstacle of \textit{ab initio} calculations employing EOM-CC theory,
and CC theory in general, is the high computational cost and excessive
memory requirements. In addition to that, CC theory
explicitly incorporates long-range electronic exchange and correlation contributions,
making convergence to the bulk-limit significantly slower than it is the case for DFAs.
Due to the
high computational cost of CC methods, it is even more challenging
than for DFA and $GW$ calculations to converge to the \gls{tdl}, 
which is approached by increasing the number of particles $N_{\text{part}}$ in the 
simulation cell, $N_{\text{part}}\to\infty$,  while keeping the particle density constant. 
The $GW$ approximation partly ameliorates this problem by adding corrections
for the long-range behavior of the dielectric function using
$k\cdot p$-perturbation theory, which are often referred to as
head- and wing-corrections~\cite{hybertsen1987ab}. 
In recent years,
studies of electronic band gaps via Quantum Monte Carlo (QMC)
methods have been published as well~\cite{hunt2018quantum, hunt2020diffusion},
where finite-size effects were also discussed as one of the major sources of
error. For these QMC band gaps, the $N$-electron ground-state
and the electronic state with one electron more ($N+1$) or less ($N-1$) were
determined separately and the energy difference of these states was computed
to obtain the band gap value. As the leading-order contribution to the
finite-size error, the interaction of the added particle ($N+1$) or
hole ($N-1$) with its periodic images was identified, which was corrected
by subtracting the screened Madelung term from the quasi-particle band gap.
Higher-order finite-size errors resulting from multipole moments of the
charged states were corrected by means of system size extrapolation.
Unfortunately, these approaches are not straight-forwardly applicable to
EOM-CC methods: The dielectric function, and therewith the head- and
wing-correction to it, is not directly accessible in the canonical
EOM-CCSD formulation. For that, linear response CC theory would be
necessary~\cite{sekino1984linear}. In contrast to the QMC ansatz,
CC methods do not use trial wavefunctions but generally rely on the
\gls{hf} wave function as a starting point, which already
incorporates the Madelung term. Instead, one needs to resort to
extrapolation techniques
or more sophisticated
finite-size error estimations using the transition structure
factor~\cite{liao2016communication, gruber2018applying}, which,
however, has been only
formulated for ground-state CC theory so far and is only a viable
technique if the correlation structure factor is represented in a 
plane wave (PW) basis.
Using a very different approach for the simulation of crystalline systems
in the CC framework, it has been
demonstrated that via  a cluster embedding approach accurate band gap
predictions can be achieved~\cite{dittmer2019accurate}. It must be stressed,
however, that the results discussed in the present work
assume periodic boundary conditions. A second point of departure is that
the work on band gaps from cluster embedding techniques utilized similarity transformed
EOM (STEOM) theory, while we in this work explore the EOM-CC method.

In this work, all CC and EOM-CC calculations were performed using a super cell approach.
Even though
it is entirely sufficient and -- due to the exploitation of the translation symmetry --
computationally significantly more efficient to calculate the band gap of a perfect crystal
on a regular $k$-grid in reciprocal space using the primitive unit cell, a $k$-point aware,
block-sparse treatment of the CC and EOM-CC equations is not yet available in CC4S.
Even for DFAs, it is well known that the super cell size convergence is an impractical 
approach for solving a proper $k$-point summation.
The IP- and EA-EOM-CCSD method exhibit a computational scaling of
$N_o^3N_v^2N_k^3$ and $N_oN_v^4N_k^3$~\cite{mcclain2017gaussian}, respectively, where $N_o$, $N_v$ and $N_k$ denote
the number of occupied and unoccupied orbitals per unit cell and the number of $k$-points.
The memory scaling is dominated by the $T_2$-amplitudes and is proportional to
$N_o^2N_v^2N_k^3$. If instead of $N_k$ $k$-points, a super cell approach with
$N_u=N_k$ unit cells is employed, a $(N_uN_o)^3(N_uN_v)^2 = N_o^3N_v^2N_u^5$
and a $(N_uN_o)(N_uN_v)^4 = N_oN_v^4N_u^5$ computational scaling for the
IP- and EA-EOM-CCSD method, respectively, is the consequence. Analogously,
the memory scaling becomes proportional to $N_o^2N_v^2N_u^4$.
Thus, a $k$-point based treatment of the EOM-CC equations
would result in a reduced computational scaling of a factor
$N_k^2$ and a reduction in memory scaling by a factor
of $N_k$. 
It must, however, be stressed that in accordance with
Bloch's theorem, the \gls{bvk} cell of a $M\times K\times L$
super cell evaluated at a single $k$-point $\bm{k}_{\text{off}}$
is identical to the \gls{bvk} cell resulting from a
primitive unit cell being evaluated on a regular $M\times K\times L$
$k$-grid shifted by $\bm{k}_{\text{off}}$. Hence, even though the super cell
approach is notably more computationally expensive, the numerical EOM-CCSD
results presented here are not affected by this choice. Still, as is well-known
for standard electronic-structure theory, only the $k$-summation approach
is practically feasible in order to achieve covergence. 

For the EOM-CC methods, one is currently forced to perform
calculations of increasing system size and perform an extrapolation
to the \gls{tdl}~\cite{mcclain2017gaussian,Gallo2021,Vo2024}, which requires knowledge
about the convergence rate of the correlation energy with respect to
the system size.  While this convergence rate has been studied for the ground-state of the
3-dimensional case of a bulk solid~\cite{xing2024inverse,gruber2018applying},
the formal convergence behavior for electronically excited states (charged or neutral) for any dimension is unknown.
\\
In this work, we formally derive the analytical expression governing the convergence
rate of the band gap on the IP/EA-EOM-CCSD level of
theory. Subsequently, we verify the correctness of the derived
expression by applying it to the band gap of a single chain of
\textit{trans}-Polyacetylene, demonstrating an efficient 
extrapolation approach to the \gls{tdl}. 
Furthermore, by repeating the
calculations using
the $G_0W_0$ method with a \gls{hf} starting point ($G_0W_0$@HF), we show that
the derived convergence rate is not specific to periodic EOM-CC theory
but can be used for other correlated methods as well.

\section{\label{sec:theory}Theory}
\subsection{\label{sec:eom-cc}The EOM-CC theory}
The EOM-CC framework is an extension to ground-state CC
theory, to compute properties of excited states.  Depending on the
nature of the excitations, different EOM-CC methods are available: The
most prominent ones are EE-EOM-CC, for neutral electronic excitations,
IP-EOM-CC for ionization processes and EA-EOM-CC for electron
attachment processes. For the present work only the latter two methods
are of importance, as the fundamental band gap of a material is given
by the difference of its lowest ionization potential and electron affinity.

The starting point of the EOM-CC method is the ground-state CC
many-electron wave function $\ket{\Psi_{0}}$, which is defined by an
exponential ansatz

\begin{eqnarray}\label{eq:exp-ansatz-CC}
    \ket{\Psi_{0}} = e^{\hat{T}}\ket{\Phi_{0}}\text{,}
\end{eqnarray}

with the Slater determinant $\ket{\Phi_0}$, which is the ground-state wave
function of a preceding mean-field calculation, usually \gls{hf}.
$\hat{T} = \sum_{n}^{M} \hat{T}_{n}$ is the so-called cluster
operator, which can excite up to $M$ electrons:

\begin{eqnarray}\label{eq:cluster-op}
    \hat{T} &&=\hat{T}_1 + \hat{T}_2 + \cdots+\hat{T}_{M}\nonumber\\
    &&=\sum_{i,a}t^{a}_{i}\hat{a}^{\dagger}_{a}\hat{a}_i + \sum_{i,j,a,b}\frac{1}{4}t^{ab}_{ij}
    \hat{a}^{\dagger}_a\hat{a}^{\dagger}_{b}
    \hat{a}_{j}\hat{a}_{i} + \cdots\nonumber\\
    &&+\left(\frac{1}{M!}\right)^2
    \sum_{\substack{i,j\cdots \\ a,b\cdots}}
    t^{ab\cdots}_{ij\cdots}
    \hat{a}^{\dagger}_{a}\hat{a}^{\dagger}_{b}\cdots
    \hat{a}_{j}\hat{a}_{i}
\end{eqnarray}

with the coefficients $t^{a}_{i}$,
$t^{ab}_{ij}$,$\cdots$ being again the
cluster amplitudes and $\hat{a}^{\dagger}_p$ and $\hat{a}_q$
the creation/annihilation operators
in second quantization, creating/annihilating an electron in orbital $p$/$q$.
The notation in Equation~\ref{eq:cluster-op} is
such that $i$,$j$,$k$
denote occupied  and
$a$,$b$,$c$ unoccupied spin orbitals. 
If $M$ is equal to the number of electrons $N$ of the system, the
ansatz in Equation~\ref{eq:exp-ansatz-CC} is exact. For reasons
of otherwise impractical computational scaling, $\hat{T}$ is truncated
in practice.
For example, $M=2$ corresponds to CC theory
with single- and double excitations (CCSD),
which is also used in this work.

To compute the wave function of the $n$-th
excited state, the EOM-CC framework
starts from a linear ansatz

\begin{eqnarray}
    \ket{\Psi_{n}} = \hat{R}_n\ket{\Psi_{0}}
\end{eqnarray}

where $\hat{R}_n$ is an excitation operator
similar to the cluster operator $\hat{T}$.
For IP-EOM-CC and EA-EOM-CC, $\hat{R}_{n}$
assumes the form

\begin{subequations}
    \begin{eqnarray}\label{eq:ip-operator}
        \hat{R}^{\text{IP}}_{n} =
        \sum_{i} r_{i,n}\hat{a}_{i} +
        \sum_{ija} r^{a}_{ij,n}\hat{a}^{\dagger}_{a}\hat{a}_{j}\hat{a}_{i} + \cdots
    \end{eqnarray}
    \begin{eqnarray}\label{eq:ea-operator}
        \hat{R}^{\text{EA}}_{n} =
        \sum_{a} r^{a}_{n}\hat{a}^{\dagger}_{a} +
        \sum_{iab} r^{ab}_{i,n}\hat{a}^{\dagger}_{a}\hat{a}^{\dagger}_{b}\hat{a}_{i} + \cdots\text{.}
    \end{eqnarray}
\end{subequations}

The $r$-coefficients contain the description of the wave function of
the $n$-th excited state and will be summarized under the excitation vectors
$|R_{n}^{\text{IP}}\rangle$ or $|R_{n}^{\text{EA}}\rangle$ .
The objective of the EOM-CC methodology
is to determine these coefficients.  In analogy
to the $\hat{T}$-operator, the operators defined in Equations~\ref{eq:ip-operator}
and \ref{eq:ea-operator} contain all possible
processes to excite an $N$-electron system into an $(N-1)$-state and
an $(N+1)$-state, respectively. However, for the same reasons stated
before for ground-state CC, $\hat{R}^{\text{IP}}_n$ and
$\hat{R}^{\text{EA}}_{n}$ are truncated in practice. The most common
approximation for EOM-CC involves only the first two terms of Equations
\ref{eq:ip-operator} and \ref{eq:ea-operator} and is termed
IP-EOM-CCSD and EA-EOM-CCSD, respectively.  Consequently, IP-EOM-CCSD
accounts only for 1-hole- and 2-hole-1-particle-excitation processes,
while EA-EOM-CCSD is restricted to 1-particle- and 2-particle-1-hole
processes. We stress that, here, the term excitation process is not restricted
to charge neutral processes but includes the removal and addition of
electrons.

In order to determine the excited states $|R_{n}^{\text{IP/EA}}\rangle$ and
the related IP- or EA-energy, the eigenproblem

\begin{subequations}
   \begin{eqnarray}
         \bar{H}|R_{n}^{\text{IP}}\rangle =
        \text{IP}_{n}|R_{n}^{\text{IP}}\rangle
    \end{eqnarray}
    \begin{eqnarray}
        \bar{H}|R_{n}^{\text{EA}}\rangle =
        \text{EA}_{n}|R_{n}^{\text{EA}}\rangle
    \end{eqnarray}
\end{subequations}

needs to be solved, where $\bar{H} = e^{-\hat{T}}\hat{H}e^{\hat{T}}$
is the similarity-transformed Hamiltonian, with 
$\hat{T}$ being the ground-state CC cluster operator.
Due to the
generally intractable size of $\bar{H}$ in the chosen representations,
the eigenvalues and -vectors cannot be determined directly but need to be computed via an
indirect approach like Davidson's method~\cite{Hirao1982}.

Once the excited states $|R_{n}^{\text{IP/EA}}\rangle$ are obtained, the
corresponding IPs or EAs can be computed via

\begin{eqnarray}\label{eq:ip-ea-expectation-val}
    \text{IP}_n/\text{EA}_n =
    \frac{\langle R_{n}^{\text{IP/EA}}|\bar{H}|R_{n}^{\text{IP/EA}}\rangle}{\langle R_{n}^{\text{IP/EA}}|R_{n}^{\text{IP/EA}}\rangle}\text{.}
\end{eqnarray}

Note here, that even though
$\bar{H}$ is for the given $\hat{T}$ non-symmetric and therefore every eigenvalue
is associated with both a left and right eigenvector,
the calculation of the eigenvalue in
Eq.~(\ref{eq:ip-ea-expectation-val}) only requires the knowledge of one of the eigenvectors.
The working equations for $\bar{H}|R_{n}^{\text{IP/EA}}\rangle$ can be
found in, \textit{e.g.}~\cite{mcclain2017gaussian}.
\\
\\
Finally, note that while the theoretical framework of CC theory for the ground
and excited states was elucidated using spin orbitals, henceforth all quantities
will be expressed in terms of (spin-independent) spatial orbitals, as all
the results shown in this work have been obtained without the consideration of
spin degrees of freedom.

\subsection{\label{sec:eom-sf}The EOM-CC structure factor}
Let us now introduce an expression for the IPs and EAs
that makes it possible to analyze their dependence on the interelectronic
distance.
A closer look of the working
equations show
that all contributions to the expectation value in
Eq.~(\ref{eq:ip-ea-expectation-val}) constitute contractions of cluster
amplitudes (see Eq.~(\ref{eq:cluster-op})), $r$-coefficients
(Eqs.~(\ref{eq:ip-operator})/(\ref{eq:ea-operator})) and Coulomb
integrals

\begin{eqnarray}\label{eq:def-coulomb-integral}
    V^{pq}_{rs} = \iint \mathrm{d}\mathbf{r}\mathrm{d}\mathbf{r}'
    \frac{\phi_{p}^*(\mathbf{r})\phi^{*}_q(\mathbf{r}')
    \phi_{r}(\mathbf{r})\phi_{s}(\mathbf{r}')}{|\mathbf{r}-\mathbf{r}'|}\text{,}
\end{eqnarray}

where $\phi_p(\mathbf{r})$ denotes a single-particle
state of the underlying mean-field
theory (\gls{hf} in this work).

By assuming \gls{bvk} boundary conditions we can introduce the co-densities in reciprocal
space
\begin{eqnarray}\label{eq:def-co-densities}
    C^{p}_{r}(\mathbf{q}) =
    \int \mathrm{d}\mathbf{r} \phi^{*}_p(\mathbf{r})\phi_{r}(\mathbf{r})e^{-i\mathbf{qr}}.
\end{eqnarray}

One can rewrite Eq.~(\ref{eq:def-coulomb-integral}) as

\begin{eqnarray}\label{eq:co-dens-decomp-V}
    V^{pq}_{rs} =\sum_{\mathbf{q}} w_{\mathbf{q}} {C^{r}_{p}}^*(\mathbf{q})v({\mathbf{q}})C^{q}_{s}(\mathbf{q})\text{,}
\end{eqnarray}

The discrete $\mathbf{q}$-vectors in Equation \ref{eq:def-co-densities} and \ref{eq:co-dens-decomp-V}
lie on a grid in reciprocal space and
$v(\mathbf{q}) = \frac{4\pi}{|\mathbf{q}|^2}$ is
the Coulomb potential in reciprocal space for the three-dimensional case.
We stress that the $\mathbf{q}$-mesh is used to represent the co-densities in Fourier space.
If the single-particle states are expressed using Bloch's theorem such that $\phi_{s}(\mathbf{r})=e^{i\mathbf{k}_s\mathbf{r}}u_{s}(\mathbf{r})$,
where $u_{s}(\mathbf{r})$ is a cell periodic function and $\mathbf{k}_s$ is a wave vector in the first Brillouin zone,
the $\mathbf{q}$-vectors correspond to the difference between the corresponding
wave vectors and a reciprocal lattice vector of the periodic unit cell.
If the single-particle states are represented using a supercell approach, the $\mathbf{q}$-vectors correspond to
reciprocal lattice vectors of the supercell. Both approaches are formally equivalent, although
Bloch's theorem enables a computationally more efficient implementation.
$w_{\mathbf{q}}$ is a weighting factor that depends on the employed integration grid and method.
It follows from Eq.~(\ref{eq:def-co-densities})
that $C^{q}_{s}(0)$  is equivalent to the
overlap between the two involved single particle states
\begin{eqnarray}\label{eq:co-dens-at-q0}
    C^{p}_{r}(\mathbf{q}=0) = \delta_{p,r}.
\end{eqnarray}

The terms contributing to the expectation value of
IP- and EA-EOM-CCSD of Equation~\ref{eq:ip-ea-expectation-val}, that is the IP and EA, can be broadly
separated into two types, single-body and many-body
contributions.

We define single-body mean-field contributions to explicitly depend on the
Fock matrix elements $f^{p}_{q}$ and only contain Coulomb integrals
implicitly by virtue of the Hartree and exchange contribution of the Fock matrix elements. A
representative single-body mean-field contribution to the
IP and EA is given by
\begin{figure}
    \centering
    \includesvg[width=0.5\textwidth]{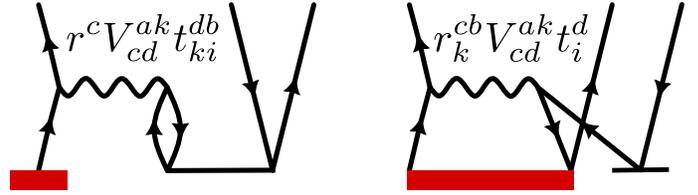}
    \caption{Goldstone diagrams and algebraic expressions for two exemplary many-body processes in the EA-EOM-CCSD equations.}
    \label{fig:1-2-body-diagrams}
\end{figure}

\begin{subequations}\label{eq:single-body-contributions}
    \begin{eqnarray}
        \text{IP}_n = \cdots -r^{ij*}_{b,n} f^{l}_{j}r^{b}_{il,n}+\cdots
    \end{eqnarray}
    and
    \begin{eqnarray}
        \text{EA}_n = \cdots +r^{j*}_{ab,n} f^{a}_{c} r^{cb}_{j,n}+\cdots\text{.}
    \end{eqnarray}
\end{subequations}

We define $\text{IP}^{(1)}_n$/$\text{EA}^{(1)}_n$ as the sum of all single-body mean-field terms
in the expression for $\text{IP}_n$/$\text{EA}_n$.
It should be noted that the commutator expansion of the similarity transformed Hamiltonian
also gives rise to effective single-body contributions originating from the contraction
of the two-body Coulomb operator with different orders of $\hat{T}_1$ and $\hat{T}_2$.
However, in this work we choose
to include only terms from the underlying mean-field Hamiltonian in our definition
of single-body terms.

It follows from our definition of $\text{IP}^{(1)}_n$/$\text{EA}^{(1)}_n$ that all remaining
contributions to $\text{IP}_n$/$\text{EA}_n$ are referred to as many-body terms and
depend explicitly on Coulomb integrals given by, for example,

\begin{subequations}\label{eq:many-body-contributions}
    \begin{eqnarray}
        \text{IP}_n = \cdots
        -2r^{i*}_n V^{kl}_{cd} t^{cd}_{il} r_{k,n}
        +\cdots
    \end{eqnarray}
    and
    \begin{eqnarray}
        \text{EA}_n = \cdots
        -2r^{*}_{a,n} V^{kl}_{cd} t^{ad}_{kl} r^{c}_n
        +\cdots\text{,}
    \end{eqnarray}
\end{subequations}

where $r^{i}_n,r_{a,n}$ and $r^{a}_{ij,n}, r^{ab}_{i,n}$
denote the single- and double excitation coefficients
of the $n$-th IP- and EA-EOM-CCSD excitation operator $|R_n^{\text{IP/EA}}\rangle$, respectively.
A diagrammatic representation of two exemplary many-body EA-EOM-CCSD contributions is shown in Figure \ref{fig:1-2-body-diagrams}.

By replacing all explicit occurrences of the Coulomb integrals $V^{pq}_{rs}$ in the many-body contributions
 by the decomposition in Equation (\ref{eq:co-dens-decomp-V})
and by contracting over all particle- and hole-indices
involved in the evaluation of Eq.~(\ref{eq:ip-ea-expectation-val}),
one arrives at an expression for
the $\text{IP}_n$ and $\text{EA}_n$ expectation value as a
sum of $\text{IP}^{(1)}_n$/$\text{EA}^{(1)}_n$ and a product
of the EOM-CC structure factor $S_{n}^{\text{IP/EA}}(\mathbf{q})$ with the Coulomb potential $v(\mathbf{q})$ as
shown in Eqs.~(\ref{eq:ip-sf}) and (\ref{eq:ea-sf})

\begin{subequations}
  \begin{eqnarray}\label{eq:ip-sf}
    \text{IP}_{n} = \sum_{\mathbf{q}}w_{\mathbf{q}} S^{\text{IP}}_{n}(\mathbf{q})v({\mathbf{q}})+\text{IP}^{(1)}_n
  \end{eqnarray}
  \begin{eqnarray}\label{eq:ea-sf}
	  \text{EA}_{n} = \sum_{\mathbf{q}}w_{\mathbf{q}} S^{\text{EA}}_{n}(\mathbf{q})v({\mathbf{q}})+\text{EA}^{(1)}_n\text{,}
  \end{eqnarray}
\end{subequations}

where the first term of Equation~\ref{eq:ip-sf} (Equation~\ref{eq:ea-sf}) contains all the many-body,
or ``correlation'' contributions to the $\text{IP}_n$ ($\text{EA}_n$) as introduced in Equation \ref{eq:many-body-contributions},
while the second term $\text{IP}^{(1)}_n$ ($\text{EA}^{(1)}_n$) denotes the sum of all single-body
mean-field contributions defined in Equation~\ref{eq:single-body-contributions}.

The above expression gives access to the dependence of the ``correlation'' energy
contribution to $\text{IP}_{n}$/$\text{EA}_{n}$ on the momentum transfer vector $\mathbf{q}$,
making it possible to perform a Fourier transform of $S^{\text{IP/EA}}_{n}(\mathbf{q})$
and attribute contributions to $\text{IP/EA}_{n}$
on the inter-electronic distance.
In this work, however, we restrict ourselves to studying the dependency on
the momentum transfer vector $\mathbf{q}$.


\subsection{The EOM-CC structure factor at $\mathbf{q}$=0}
\label{sec:eom-sf-at-0}
Since $C^{q}_{s}(0)=\delta_{q,s}$, the contributions to
$S^{\text{IP/EA}}_n(\mathbf{q}=\mathbf{0})$ can be further simplified.
In particular, one discovers that
$S^{\text{IP/EA}}_n(\mathbf{q}=\mathbf{0})$ corresponds to the many-body character of the
IP or EA state. We define
the single-body character $p_1$ of an IP or EA state as

\begin{subequations}
    \begin{eqnarray}
        p_{1,n} = \sum_{i} |r_{i,n}|^2
    \end{eqnarray}
    and
    \begin{eqnarray}
        p_{1,n} = \sum_{a} |r^{a}_{n}|^2\text{,}
    \end{eqnarray}
\end{subequations}

respectively, where the $r$-coefficients
from Equation \ref{eq:ip-operator} and \ref{eq:ea-operator} are
used. In this definition, the single-body character $p_{1,n}$ of the
$n$-th IP or EA state is given by the contribution of single-hole or
single-particle processes to the overall description of the $(N+1)$/$(N-1)$ wave
function. For a normalized excited state $|R_{n}^{\text{IP/EA}}\rangle$,
$p_{1,n}$ lies between 0 and 1.
As we will show, for both IP- and EA-EOM-CCSD,
\begin{eqnarray}\label{eq:s0-is-manybody-character}
    S^{\text{IP/EA}}_{n}(\mathbf{q}=\mathbf{0}) =
    -(1-p_{1,n})
\end{eqnarray}
holds. Thus, the value of the
EOM-CC structure factor at $\mathbf{q}=\mathbf{0}$ is equal to the negative
many-body character of the excitation.

This results from Equation \ref{eq:co-dens-at-q0}, so that the only terms of the EOM-CCSD
equations, that contribute to the value
of the EOM-CC structure factor at $\mathbf{q}=0$ are those
which feature Coulomb integrals of type $V^{ab}_{cd}$,
$V^{ij}_{kl}$, $V^{ai}_{bj}$ or $V^{ia}_{jb}$. That is,
the integrals must be representable as products of hole-hole or particle-particle co-densities following Equation
\ref{eq:co-dens-decomp-V}. In the case of IP-EOM-CCSD,
there are 5 terms which contribute to $S(\mathbf{q}=0)$, which are

\begin{equation}\label{eq:ip-s0-contributions}
\begin{split}
    S^{\text{IP}}_{n}(\mathbf{q}=0)=
    \Bigg [ \Bigg .-&r^{ij*}_{b} C^{l*}_{j}(\mathbf{q}) C^{b}_{d}(\mathbf{q})r^{d}_{il}\\
    -&r^{ij*}_{b} C^{k*}_{i}(\mathbf{q}) C^{b}_{d}(\mathbf{q})r^{d}_{kj}\\
    &r^{ij*}_{b} C^{k*}_{i}(\mathbf{q}) C^{l}_{j}(\mathbf{q})r^{b}_{kl}\\
    -&r^{ij*}_{b} C^{k*}_{i}(\mathbf{q})C^{b}_{d}(\mathbf{q}) t^{d}_{j} r_{k}\\
    &r^{ij*}_{b} C^{k*}_{i}(\mathbf{q}) C^{l}_{j}(\mathbf{q}) t^{b}_{l} r_{k}
    \Bigg. \Bigg]_{\mathbf{q}=0}\text{.}
\end{split}
\end{equation}

By making use of the fact that at $\mathbf{q}=0$, the co-densities reduce to overlap integrals between the single-particle states (see Equation \ref{eq:co-dens-at-q0}),
Equation \ref{eq:ip-s0-contributions} reduces to

\begin{align*}
S^{\text{IP}}_{n}(\mathbf{q}=0)=-&r^{ij*}_{b}r^{b}_{ij}
    -r^{ij*}_{b} r^{b}_{ij}
    +r^{ij*}_{b} r^{b}_{ij}\\
    -&r^{ij*}_{b}t^{b}_{j} r_{i}
    +r^{ij*}_{b} t^{b}_{j} r_{i}\\
    =-&r^{ij*}_{b}r^{b}_{ij}\text{,}
\end{align*}

which -- for a normalized EOM-CCSD excitation vector --
is equivalent to Equation \ref{eq:s0-is-manybody-character}. The derivation for the EA-EOM-CCSD structure factor is analogous.

We emphasize that the value for $S^{\text{IP/EA}}_{n}(\mathbf{q}=\mathbf{0})$ is a direct
consequence of the chosen definition of $\text{IP}^{(1)}_n$/$\text{EA}^{(1)}_n$
and the eigenvalue in Eq.~\ref{eq:ip-ea-expectation-val}. Although this choice is
helpful for the analysis of finite-size errors, there also exist
alternative approaches, for example, using left eigenvectors as bra-states in
Eq.~\ref{eq:ip-ea-expectation-val}, resulting in a different value for $p_{1,n}$
and $S^{\text{IP/EA}}_{n}(\mathbf{q}=\mathbf{0})$.
We expect that for the systems studied in this work, which exhibit a very small
many-body character, both approaches discussed above would yield very similar results.

\subsection{\label{sec:eom-sf-long-range}Long-range behavior of the EOM-CC structure factor}
To determine the asymptotic behavior of the
EOM-CC structure factor in the long-wavelength limit, that is for $|\mathbf{q}|\to 0$,
one can perform a Taylor series expansion around $\mathbf{q}=0$ by computing the
derivatives
of the EOM-CC structure factor with respect to $\mathbf{q}$.
The only explicit $\mathbf{q}$-dependence of that quantity comes from
the co-densities $C^{p}_{r}(\mathbf{q})$.
As laid out in Section
\ref{sec:eom-sf}, $v(\mathbf{q})$ is not included in the expression for
$S^{\text{IP/EA}}(\mathbf{q})$, so that the products of co-densities
$C^{p*}_{r}(\mathbf{q})C^{q}_{s}(\mathbf{q})$ are the only
$\mathbf{q}$-dependent quantities. We note that
we neglect the implicit $\mathbf{q}$ dependence possibly introduced by the EOM-CC amplitudes
($r_i$, $r^a$, $r_{ij}^{a}$, $r_{i}^{ab}$). This assumed $\mathbf{q}$ independence of the 
$r$-amplitudes for $\mathbf{q}\to 0$ is justified by the fact that for systems with a band gap none of the quantities 
that appear in the EOM-CC equations, which are the Fock matrix elements, the Coulomb integrals 
and the ground-state $T$-amplitudes, depend on $\mathbf{q}$ (assuming the crystal momentum is conserved).
While this is evident for the Fock matrix and the Coulomb integrals, the $\mathbf{q}$ independence
of the $T_2$-amplitudes is necessary for the ground-state transition 
structure factor to yield the correct $1/N_k$ convergence behavior
in the thermodynamic limit ($\mathbf{q}\to 0$)~\cite{liao2016communication}. For the $T_1$ amplitudes,
which do not explicitely appear in the expression of the ground-state CC correlation energy or the
transition structure factor, a similiar argument as for the EOM-CC equations applies: Since the
$T_1$-amplitudes are determined iteratively from $\mathbf{q}$ independent quantities, it is 
reasonable to assume that they themselves do not exhibit any dependence on $\mathbf{q}$ either. 

By making use of the product rule and by realizing
that a co-density at $\mathbf{q}=0$ is equivalent to the
overlap between the two involved single particle states as defined by Eq.~(\ref{eq:co-dens-at-q0}),
one finds that

\begin{eqnarray}\label{eq:co-dens-1st-deriv}
    \frac{\partial}{\partial \mathbf{q}} C^{p*}_{r}(\mathbf{q})C^{q}_{s}(\mathbf{q})\bigg|_{\mathbf{q}=0} =
    \delta_{p,r} \frac{\partial}{\partial \mathbf{q}} C^{q}_{s}(\mathbf{q})\bigg|_{\mathbf{q}=0} +\nonumber\\
    \delta_{q,s} \frac{\partial}{\partial \mathbf{q}} C^{p*}_{r}(\mathbf{q})\bigg|_{\mathbf{q}=0}
\end{eqnarray}

Eq.~(\ref{eq:co-dens-1st-deriv}) reveals, that
if the co-densities and by extension the Coulomb integrals in the
EOM-CCSD equations would only couple holes with particles, the first
derivative of the EOM-CC structure factor would vanish.
In passing we note that this is the situation
for the ground-state CC transition structure factor, which is why the
lowest $\mathbf{q}$-order in the long-wavelength limit for insulators is
quadratic~\cite{liao2016communication}.  The EOM-CC working equations,
however, do also feature contractions with Coulomb integrals, which
couple holes with holes and particles with particles, so that formally
a linear contribution in $\mathbf{q}$ to the EOM-CC structure factor must be considered,
because in general
$\frac{\partial}{\partial \mathbf{q}} C^{q}_{s}(\mathbf{q})\bigg|_{\mathbf{q}=0}\neq 0$.

By repeating this procedure for the second derivative,
one finds that the quadratic contribution
does in general not vanish either, so that we can approximate
the asymptotic behavior of the EOM-CC
structure factor up to second order

\begin{eqnarray}\label{eq:eom-sf-asymptotic}
    \lim_{|\mathbf{q}|\to 0}\, S^{\text{IP/EA}}(|\mathbf{q}|) \approx
    S^{\text{IP/EA}}_{n}(\mathbf{q}=0) + \alpha \cdot|\mathbf{q}| + \beta \cdot|\mathbf{q}|^2\text{,}
\end{eqnarray}

where $\alpha$ and $\beta$ are constants and $S^{\text{IP/EA}}_{n}(\mathbf{q}=0)$ is the value
of the EOM-CC structure factor at $\mathbf{q}=0$ as discussed in Section~\ref{sec:eom-sf-at-0}.
We note that in Equation~\ref{eq:eom-sf-asymptotic},
we assume $S^{\text{IP/EA}}_{n}(\mathbf{q})$ to be spherically symmetric, which -- while not generally the case -- simplifies the 
following derivation and analysis.

Note, that the above derivation -- while based on some assumptions-- did not 
depend in any way on the dimensionality of the system.

\subsection{Convergence rate of IPs, EAs and band gaps
to the TDL}\label{sec:convergence-rate-derivation}

Based on the discussion from the previous sections, we now want to make a statement about the asymptotic
convergence behavior of computed IPs, EAs and band gaps to the \gls{tdl}.
The \gls{tdl} is approached for supercells with increasing size or increasingly dense $k$-mesh used to sample
the first Brillouin zone.
This corresponds to a continuous  $\mathbf{q}$-representation
of the EOM-CC structure factor such that Eqs.~(\ref{eq:ip-sf}) and
(\ref{eq:ea-sf}) are given by

\begin{eqnarray}\label{eq:ip-ea-sf-integral}
	\text{IP}_{n}/\text{EA}_{n} = \int {\mathrm{d}\mathbf{q}}\; S^{\text{IP/EA}}_{n}(\mathbf{q})v({\mathbf{q}}) + \text{IP}_{n}^{(1)}/\text{EA}_{n}^{(1)}\text{,}
\end{eqnarray}

Finite size errors of, for example, IP/EA energies are defined by the difference between calculations
using finite system sizes and the \gls{tdl}.
For the above equation this corresponds to the difference between a continuous integration and a
discrete sampling. Similar to the case of ground state CC calculations, we assume that the
values of the EOM-CC structure factor at the sampled $\mathbf{q}$ points converge rapidly with 
respect to the employed system size, which will be later verified numerically.
In other words, $S^{\text{IP/EA}}_{n}(\mathbf{q})=S^{\text{IP/EA-(TDL)}}_{n}(\mathbf{q})$,
where $S^{\text{IP/EA}}_{n}(\mathbf{q})$ and $S^{\text{IP/EA-(TDL)}}_{n}(\mathbf{q})$ refer to the structure
factor from the finite system and \gls{tdl}, respectively. Note that in this approach $\mathbf{q}$ is restricted to a
discrete subset depending on  the system size, whereas its continuous in the \gls{tdl}.
In this case, the largest contribution to the finite size error originates from the
employed finite simulation cell size and the neglect of long-range interactions in real space
corresponding to short $\mathbf{q}$-vectors.
Under these assumptions it is reasonable to define the following estimate for the finite size
error of the correlation contribution $\Delta_{\text{FS}}^{\text{IP/EA}}$

\begin{eqnarray}\label{eq:eom-fs-integral}
    \Delta_{\text{FS}}^{\text{IP/EA}} =
    \int_{\Omega_{q_\text{min}}} {\mathrm{d}\mathbf{q}}\;S^{\text{IP/EA}}_{n}(\mathbf{q})v({\mathbf{q}})\text{.}
\end{eqnarray}

The integral in Equation \ref{eq:eom-fs-integral} is carried out over
a sphere $\Omega_{q_\text{min}}$ centered at the $\Gamma$ point with radius $q_{\text{min}}$. The radius
should be understood as a measure for the shortest reciprocal lattice vector of
the considered simulation cell.
For the 3-dimensional case, we evaluate the integral
in Equation \ref{eq:eom-fs-integral} using the Fourier
transform of the Coulomb potential in three dimensions which is

\begin{eqnarray}
    v(\mathbf{q}) = \frac{4\pi}{\mathbf{q}^2},
\end{eqnarray}

so that -- using $q=|\mathbf{q}|$ -- a measure of the finite-size error is given by

\begin{align}\label{eq:derivation-fs-convergence-3D}
    \Delta_{\text{FS}}^{\text{IP/EA}} & \propto
    \int_{\Omega_{q_\text{min}}}
    \mathrm{d}\mathbf{q}\;S^{\text{IP/EA}}_{n}(\mathbf{q})v(\mathbf{q})\nonumber\\
    &\propto \int_{0}^{q_{\text{min}}} \mathrm{d} q\; 4\pi q^2
    S^{\text{IP/EA}}_{n}(\mathbf{q})v(\mathbf{q}) \nonumber\\
    &\propto\int_{0}^{q_{\text{min}}}
    \mathrm{d} {q}\; S^{\text{IP/EA}}_{n}(\mathbf{q}=0) + \alpha \cdot |\mathbf{q}|+ \beta \cdot |\mathbf{q}|^2 \nonumber\\
    &\propto 
    A\cdot q_{\text{min}} +
    B\cdot q^2_{\text{min}} +
    C\cdot q^3_{\text{min}}\text{,}
\end{align}

where all constants resulting from the integration
are collected in the parameters $A$, $B$ and $C$. By assuming spherical symmetry of the
EOM-CC structure factor, the derivation in Equation \ref{eq:derivation-fs-convergence-3D}
is simplified substantially. 


More practical is Equation \ref{eq:derivation-fs-convergence-3D} if expressed as a function
of the total number of $\mathbf{k}$-points $N_k$ of a uniform $k$-mesh.
Noting, that

\begin{eqnarray}
    q_{\text{min}} \propto N_k^{-\frac{1}{d}}\text{,}
\end{eqnarray}

where $d$ is the dimension of the system. By applying this
equality for the three-dimensional case, Equation \ref{eq:derivation-fs-convergence-3D} becomes

\begin{eqnarray}\label{eq:3d-convergence-rate-Nk}
    \Delta_{\text{FS}}^{\text{IP/EA}} &\propto
    A\cdot N_k^{-1/3} +
    B\cdot N_k^{-2/3} +
    C\cdot N_k^{-1}\text{.}
\end{eqnarray}

\subsubsection{Convergence rates for low dimensional systems}

Since we assume that the asymptotic behavior of the EOM-CC structure factor is
independent of the dimension of the system, we can now make use of
Equation~\ref{eq:eom-sf-asymptotic} to derive the leading-order
rate of convergence to the \gls{tdl} for one- and two-dimensional systems.
To this end we make use of the
Fourier transforms of the Coulomb potential in the two- and one-dimensional case~\cite{mihaila2011lindhard}:

\begin{eqnarray}
    v^{\text{2D}} = \frac{2\pi}{|\mathbf{q}|}
\end{eqnarray}

and

\begin{eqnarray}
    v^{\text{1D}} = -2\gamma_E + \ln{\left(\frac{1}{\mathbf{q}^2}\right)}\text{,}
\end{eqnarray}

respectively, where $\gamma_E$ denotes the Euler constant. 
Note that we employ here the lower-dimensional Fourier transforms
of the Coulomb potential even though realistic one-dimensional systems 
generally have a finite extent in the directions perpendicular to
the a one- or two-dimensional material.
In the long-wavelength limit, however, which is the focus of the present work,
these contributions become negligible in comparison to those of the
sheet or chain direction(s). This justifies the representation
of the Coulomb potential by its lower-dimensional Fourier transforms.
By repeating the same steps as for 
the derivation of the 3-dimensional convergence rate, 
we find that the finite-size error in a two-dimensional
system converges like

\begin{eqnarray}\label{eq:2d-convergence-rate-Nk}
    \Delta_{\text{FS}}^{\text{IP/EA}} &\propto
    A\cdot N_k^{-1/2} +
    B\cdot N_k^{-1} +
    C\cdot N_k^{-3/2}\text{.}
\end{eqnarray}

and for a one-dimensional system the convergence rate is
given by

\begin{equation}\label{eq:1d-convergence-rate-Nk}
  \begin{matrix}
    \Delta_{\text{FS}}^{\text{IP/EA}}(N_k) & \propto
    & A\cdot N_k^{-1}\ln{\left(N_k^{-1}\right)} + B\cdot N_k^{-1} \\
    & + & C\cdot N_k^{-2}\ln{\left(N_k^{-1}\right)} + D\cdot N_k^{-2} \\
    & + & E\cdot N_k^{3}\ln{\left(N_k^{-1}\right)} +
    F\cdot N_k^{-3}\text{.}
  \end{matrix}
\end{equation}

A summary of the derived convergence rates for
1, 2 and 3 dimensions is given in Table~\ref{tab:conv-rates-1d-to-3d}.
We reiterate that these convergence rates are estimated from the contributions
to $\text{IP}_{n}$/$\text{EA}_{n}$ around $\mathbf{q}=0$ in a sphere
with a radius decreasing as $N_k$ increases.
It should also be noted that the actual convergence rate of numerically
computed $\text{IP}_{n}$/$\text{EA}_{n}$'s depends on the chosen
treatment of the Coulomb singularity in the respective computer
implementation, which will be discussed in the following section.

\begin{table}[]
    \centering
    \caption{Convergence rates for one-, two- and three-dimensional systems up to second order.}
    \begin{tabular}{cc}
      \toprule
      Dimension &  Convergence rate as a function of $N_k$\\
      \midrule
      1D & $
           \begin{matrix}
             & A\cdot N_k^{-1}\ln{\left(N_k^{-1}\right)} + B\cdot N_k^{-1} \\
              + & C\cdot N_k^{-2}\ln{\left(N_k^{-1}\right)} + D\cdot N_k^{-2} \\
             + & E\cdot N_k^{-3}\ln{\left(N_k^{-1}\right)} +
             F\cdot N_k^{-3}
           \end{matrix}$\\
      \midrule
      2D & $A\cdot N_k^{-\frac{1}{2}} + B\cdot N_k^{-1} + C\cdot N_k^{-\frac{3}{2}}$\\
      \midrule
    3D &  $A\cdot N_k^{-\frac{1}{3}} + B\cdot N_k^{-\frac{2}{3}} + C\cdot N_k^{-1}$\\
      \bottomrule
    \end{tabular}
    \label{tab:conv-rates-1d-to-3d}
\end{table}


\subsection{Convergence rates and treatment of Coulomb singularity}

The convergence rates derived in the previous section
assume a spherically truncated integration around $\mathbf{q}=0$ to estimate
the finite size errors $\Delta_{\text{FS}}^{\text{IP/EA}}$.
In practical \textit{ab initio} calculations, however, there exist a variety
of treatments to approximate the integral around the Coulomb singularity at $\mathbf{q}=0$,
which strongly influence the convergence rates as can already be observed for the exchange energy 
contribution~\cite{Sundararaman2013,Spencer2008,Carrier2007,Gygi1986,schafer2024sampling,Irmler2018}.
We now briefly discuss the significance of  the treatment of the Coulomb singularity for
the convergence rates derived in the previous section.

For the present study, we employ a Coulomb singularity treatment
that captures the contribution of the
$S^{\text{IP/EA}}_{n}(\mathbf{q}=0)$ term to the
integral in Eq.~\ref{eq:derivation-fs-convergence-3D} exactly if
the value of $S^{\text{IP/EA}}_{n}(\mathbf{q}=0)$ is converged with respect to system size.
As a consequence, the contributions to the finite size errors proportionate to $A$ given
in Table~\ref{tab:conv-rates-1d-to-3d} will already be accounted for and the expected
next-leading order contribution to the  finite size error will be proportionate to $B$.
In particular, the plane wave basis set calculations of the present work compute
the average Coulomb kernel for the volume element at $\mathbf{q}=0$ to estimate
its contribution to the integral~\cite{schafer2024sampling}.

We note that there also exist other approaches in the literature that disregard the
Coulomb singularity contribution
and obtain EOM-CC band gaps by extrapolation.
Refs.~\cite{mcclain2017gaussian,gao2020electronic,Vo2024}
disregard the $\mathbf{q}=0$ contribution already in the underlying \gls{hf} calculation.
As a consequence the \gls{hf} band gap is underestimated
and converges only as $N_k^{-1/3}$ in three-dimensional systems.
The same applies to the finite-size error in EOM-CC theory as shown in the previous section.
It should be noted that the finite size errors from \gls{hf} and EOM-CC partly cancel each other.
However, the extrapolation procedure still requires a careful checking and can lead to errors
that are difficult to control due to next-leading order contributions to the finite size errors.

\section{Computational details}\label{sec:computational-details}
As the practical representation of the EOM-CC structure factor requires the utilization
of a plane wave basis, we employed the
Vienna Ab Initio Simulation Package (VASP)~\cite{kresse1994norm,kresse1996efficient}
in combination with the CC4S software package, a periodic CC
code~\cite{gruber2018applying}, where the working equations of the IP- and
EA-EOM-CCSD methods and their respective structure factors were implemented.
In the case of the LiH EOM-CC and ground-state CC structure factors, which are discussed in 
Section \ref{sec:lih}, a relatively small basis set with 3 virtual orbitals per occupied 
orbital was employed in order to compute the huge super cell sizes necessary to resolve 
the structure factor for very small momentum vectors.
However, we are concerned with the qualitative long-range description of the
EOM-CCSD structure factor, for which such a small plane-wave basis set is sufficient.
Similarly, the EOM-CC structure factors
for \textit{trans}-Polyacetylene (tPA) were computed using a reduced basis set of 4 virtual orbitals
per occupied orbital. Due to substantial amount of vacuum in the simulation cell of the tPA chain,
it is not practical to converge the results with respect to the plane-wave energy cut-off, which is why
VASP was only used to obtain the EOM-CC structure factor but not the electronic band gaps themselves.
Specifically for the LiH calculations, the \texttt{Li\_GW} and \texttt{H\_GW} POTCARs were employed, using
a plane-wave energy cut-off of $300\,$ eV. The tPA EOM-CCSD structure factor was computed with the \texttt{C\_GW\_new} and \texttt{H\_GW} POTCARs and an energy cut-off of $300\,$ eV.

To alleviate convergence
problems of the \gls{hf} and the CC calculations associated with
the discretization of the Brillouin zone (BZ) for the highly anisotropic simulation cells of tPA, we employ the recently
developed improved sampling method of the Coulomb potential
in VASP~\cite{schafer2024sampling}.
\\
The numerical convergence tests for the band gaps of tPA
were performed using FHI-aims~\cite{blum2009ab},
which employs numeric atomic orbitals and which has been interfaced to CC4S as well~\cite{moerman2022interface}.
To determine the convergence to the \gls{tdl}, the unit cell of a single tPA chain with 2 carbon atoms and 2 hydrogen atoms
with at least $80\,\mathring{A}$ of vacuum in each direction perpendicular to the chain was optimized
employing the B3LYP exchange-correlation
functional~\cite{becke1992density,lee1988development}, since this functional was shown 
in previous studies~\cite{hirata1998density} to reasonably reproduce the bond length alternation of tPA 
observed in experiment~\cite{yannoni1983molecular}. 
For the geometry optimization a \textit{tight} 
tier-2 basis set and a $1\times 1\times 20$ $\mathbf{k}$-grid was used. 
Due to the computational overhead associated with additional vacuum
when computed using a PW basis, the vacuum around the tPA chain was reduced to at least $7\,\mathring{A}$
for calculations with VASP.
The CC and $GW$ calculations involving FHI-aims were performed using the loc-NAO-VCC-$n$Z basis sets
developed by Zhang et al.~\cite{zhang2013numeric, zhang2019main}.
The basis set convergence for the band gap of tPA was
investigated on the EOM-CCSD level of theory, the results of which are shown in Table
\ref{tab:aims-basis-set-convergence}.

\begin{table}[h]
  \centering
  \begin{tabular}{cccc}
    \toprule
    & 2Z & 3Z & 4Z\\
    \midrule
    $E_{\text{gap}}^{1\times 1\times 6}\;$   & 5.859& 5.765& 5.702\\
    $E_{\text{gap}}^{1\times 1\times 8}\;$   & 5.403& 5.318& 5.262\\
    \bottomrule
  \end{tabular}
  \caption{Basis set convergence of the tPA band gap
    using the loc-NAO-VCC-$n$Z ($n=2,3,4$) basis set 
    with a $1\times 1\times 6$ and a  $1\times 1\times 8$
    $\mathbf{k}$-mesh. Energies are given in units of eV.}
  \label{tab:aims-basis-set-convergence}
\end{table}

As Table \ref{tab:aims-basis-set-convergence} conveys, the loc-NAO-VCC-2Z basis set allows to converge the band gap to
 $157\,\text{meV}$ for a $1\times 1\times 6$ $\mathbf{k}$-mesh and $141\,\text{meV}$ for a $1\times 1\times 8$ $\mathbf{k}$-mesh, essentially independently of the size of the BvK cell, which we consider to be sufficient for the present application as the finite-size error rather than the basis-set incompleteness error is at the center of this study. Hence, all calculations
involving FHI-aims will be performed using the loc-NAO-VCC-2Z basis set. 

In order to approximate the contribution from the singularity of the Coulomb potential,
a variation of the spherical truncation approach pioneered by Spencer and Alavi~\cite{Spencer2008} is employed
in FHI-aims. In this \textit{cut-Coulomb} approximation the long-ranged part of the Coulomb potential 
is made to decay fast to 0 beyond a set radius $r_{\text{cut}}$  by multiplying the original $\frac{1}{r}$
potential with a complementary error function~\cite{levchenko2015hybrid}

\begin{equation}
	v^{\text{cut-Coulomb}} = 
	\frac{1}{2|\mathbf{r}|}\text{erfc}[\eta(|\mathbf{r}| - r_{\text{cut}})]
\end{equation}

with $\eta$ as the inverse decay width.

\section{Results}\label{sec:results}
To determine the validity of the convergence rate of the finite-size error
for IP- and EA-EOM-CCSD, the findings will be presented as follows:
First, the fundamental properties of the EOM-CC structure factor that have been
mentioned in part already in Section \ref{sec:theory}, will be elucidated
by looking at the structure factor of the 3D LiH primitive cell repeated
periodically in one direction. Following that qualitative discussion,
the previously derived convergence rates of the EOM-CC band gap finite-size error will
be applied to a one-dimensional system, a chain of \textit{trans}-Polyacetylene. Finally, the 
analytical extrapolation expression will
be applied to the finite-size convergence of
the $G_0W_0$ band gap, testing its validity outside of EOM-CC theory.

\subsection{The EOM-CC structure factor of LiH in one dimension}\label{sec:lih}

In order to relate the herein newly derived EOM-CC structure factor to the
ground-state analogue, the 
\textit{transition structure factor}~\cite{liao2016communication, gruber2018applying}, we
want to start by computing these two quantities for the LiH chain 
using supercells, which consist of increasingly many unit cells in a single direction.
Since the purpose of this section is 
a direct comparison of the correlation structure factors of the two methods, only the $\Gamma$-point
was sampled in reciprocal space.
As noted in Section \ref{sec:eom-sf-long-range}, the derived behavior
of $S^{\text{IP/EA}}(q)$ for $q\to 0$ (and equally so for the
transition structure factor) should hold independently of the
system's dimension.

\begin{figure}
  \includegraphics[scale=0.55]{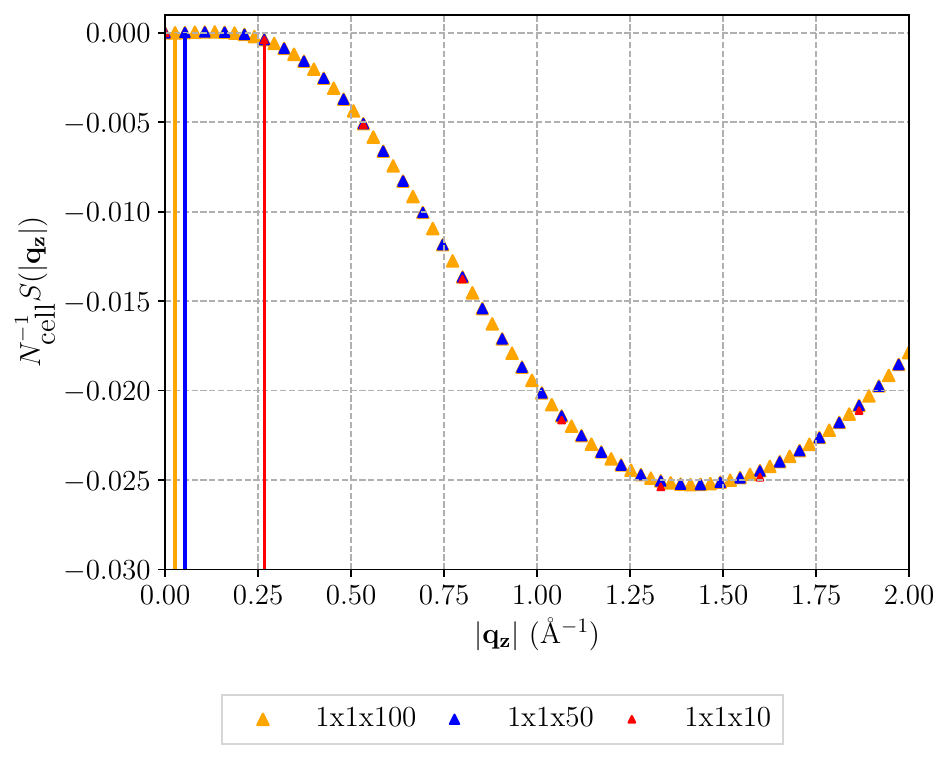}
  \caption{\label{fig:lih-sf-convergence-ground-state} Ground state CCSD transition structure factor
  of LiH plotted along the z-axis using a $1\times 1\times 10$, $1\times 1\times 50$ and 
  a $1\times 1\times 100$ supercell 
  extended in one dimension.
  The vertical lines mark the minimum distance
  between any two lattice points in the reciprocal lattice. To make the transition structure factor 
  of different BvK cell sizes comparable, the structure factor is scaled by one over the number of 
  unit cells in the respective super cell, $N_{\text{cell}}^{-1}$.}
\end{figure}

We shall start by reviewing the properties of the ground-state CCSD transition
structure factor, which is shown in Figure \ref{fig:lih-sf-convergence-ground-state} for
the LiH chain for super cell sizes of  $1\times 1\times 10$,
$1\times 1\times 50$ and $1\times 1\times 100$. Subsequently, a comparison to the EOM-CC
structure factor will be drawn.

In analogy to the EOM-CC structure factor, which upon contraction with the Coulomb potential
and integration over the reciprocal space yields 
the correlation contribution to the charged quasi-particle energies (see Equation \ref{eq:ip-ea-sf-integral}), 
the ground-state correlation energy $E^{\text{corr}}_0$ can be obtained in the same way by means 
of the transition structure factor $S(\mathbf{q})$

\begin{equation}
	E^{\text{corr}}_0 = \int {\mathrm{d}\mathbf{q}}\; S(\mathbf{q})v({\mathbf{q}})\text{.}
\end{equation}

Figure \ref{fig:lih-sf-convergence-ground-state} shows the medium- to long-range portion of
the transition structure factor in the direction of the chain, that is the $z$-direction. Even though,
$S(\mathbf{q})$ can be computed for longer $\mathbf{q}$-vectors, corresponding to short-range
processes in real space, this is not relevant to the present discussion of the finite-size convergence.
The vertical lines in Figure \ref{fig:lih-sf-convergence-ground-state} show the magnitude of the smallest $\mathbf{q}$-vector
that can be resolved in the respective super cell. That minimal $\mathbf{q}$-vector is given by the minimal distance
of two $\mathbf{k}$-points or (in a super cell formulation) by the smallest reciprocal lattice vector and corresponds
to the biggest real-space distance captured by the \gls{bvk} cell. As can be observed in Figure \ref{fig:lih-sf-convergence-ground-state},
the bigger the super cell size of LiH, the smaller does this minimal $\mathbf{q}$-vector become and
the more long-range correlation information
does the transition structure factor contain.
The transition structure factor in Figure \ref{fig:lih-sf-convergence-ground-state} features
a minimum, at some material- and state-specific distance in reciprocal space.
This distance can be interpreted as a characteristic distance for the electronic correlation, as 
we expect the
contribution of the CCSD transition structure factor to the correlation energy to be maximal in the vicinity 
of its extremum.
In the specific case of the LiH chain, one finds that the minimum is at $~1.4\;\mathrm{\AA}^{-1}$ for 
the ground state correlation energy, which corresponds to approximately a $1\times 1\times 2$ supercell.
Another fundamental property of the transition structure factor of the ground-state is the fact
that for $|\mathbf{q}|\to0$, $S(\mathbf{q})\to0$ with zero slope as is apparent in 
Figure \ref{fig:lih-sf-convergence-ground-state}.
This directly results from the sum-rule of the pair correlation function, which is the Fourier transform
of the transition structure factor.

\begin{figure}
  \includegraphics[scale=0.55]{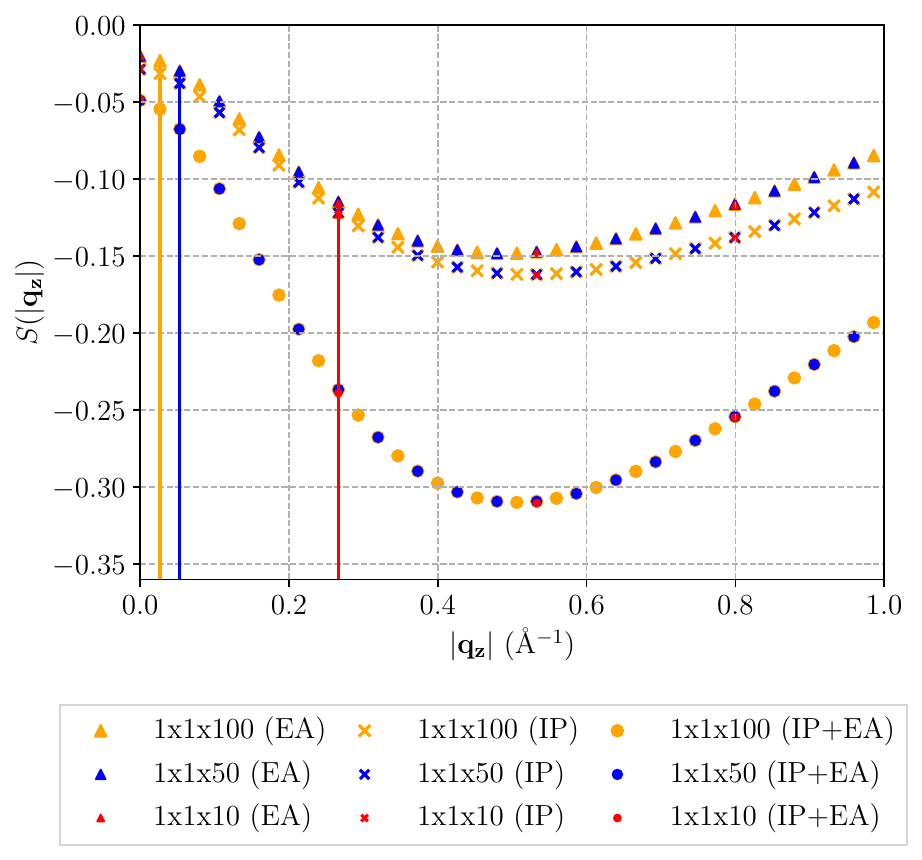}
  \caption{\label{fig:lih-sf-convergence} IP-EOM-CCSD,  EA-EOM-CCSD transition structure factors and the sum of both corresponding to the band gap
  of LiH plotted along the z-axis using a $1\times 1\times 10$, $1\times 1\times 50$ and a $1\times 1\times 100$ supercell extended in one dimension.
  The vertical lines mark the minimum distance
  between any two lattice points in the reciprocal lattice.}
\end{figure}

The EOM-CC structure factor for charged excitations exhibits qualitative differences to the ground-state
transition structure factor. The structure factor for the first IP and EA in the EOM-CCSD framework for the
LiH chain is shown in Figure \ref{fig:lih-sf-convergence}.

The EOM-CC structure factors feature a minimum as well, but it appears at 
a significantly smaller $|\mathbf{q}|$ value of $~0.5\;\mathrm{\AA}^{-1}$, 
which corresponds approximately to a $1\times 1\times 6$ super cell. This reflects the
longer range of correlation effects in the EOM-CC case in comparison to the ground-state.
This comparison
clarifies that the problem of reaching the \gls{tdl} for charged quasi-particle energies is substantially
more difficult than for the ground state case.
Another point of departure between ground state CC
and EOM-CC theory, is the value of the structure factor at $\mathbf{q}=0$. 
While in the case of the ground state the $S(\mathbf{q}=0)$ value is always 0,
in the case of EOM-CC we have previously derived that $S(\mathbf{q}=0)$ is a finite negative value in our
formulation corresponding to the many-body character of the respective excitation.

As Figure \ref{fig:lih-sf-convergence} demonstrates, all the properties discussed for the EOM-CC structure factor apply in a similar fashion to both the IP- and the EA-EOM-CCSD
case and the IP+EA case.
Let us stress once more that the IP+EA case corresponds to the electronic band gap in the presently used convention. It should be stressed at this point, that the EOM-CC structure factor that we have introduced
in this work only captures the correlation contribution to the IP and EA quasi-particle energies and to
the band gap, which are given by the many-body contributions of the EOM-CC equations as discussed
in Section \ref{sec:eom-sf}. In order to obtain the full excitation energy, we compute and converge
the generally sizable single-body contributions (compare Equation \ref{eq:single-body-contributions}) to the EOM-CCSD expectation value separately.

With respect to the previously derived asymptotic behavior of the EOM-CC structure factor for $\mathbf{q}\to 0$, it becomes immediately apparent from the LiH EOM-CC structure factors, particularly for the $1\times 1\times
100$ cell, that it does not seem to exhibit significant linear behavior at $\mathbf{q}=0$.
Furthermore, $S(\mathbf{q}=0)$ is relatively small.
Therefore, one can conclude for the anisotropic LiH super cell,
that both the constant $|\mathbf{q}|^0$ and the linear contribution $|\mathbf{q}|^{1}$ contribution of the EOM-CC structure
factor to $\Delta_{\text{FS}}^{\text{IP/EA}}$ is negligible.

\subsection{The \textit{trans}-Polyacetylene chain}\label{sec:tPA}

\subsubsection{The EOM-CC structure factor of tPA}

\begin{figure}
    \centering
    \includegraphics[scale=0.55]{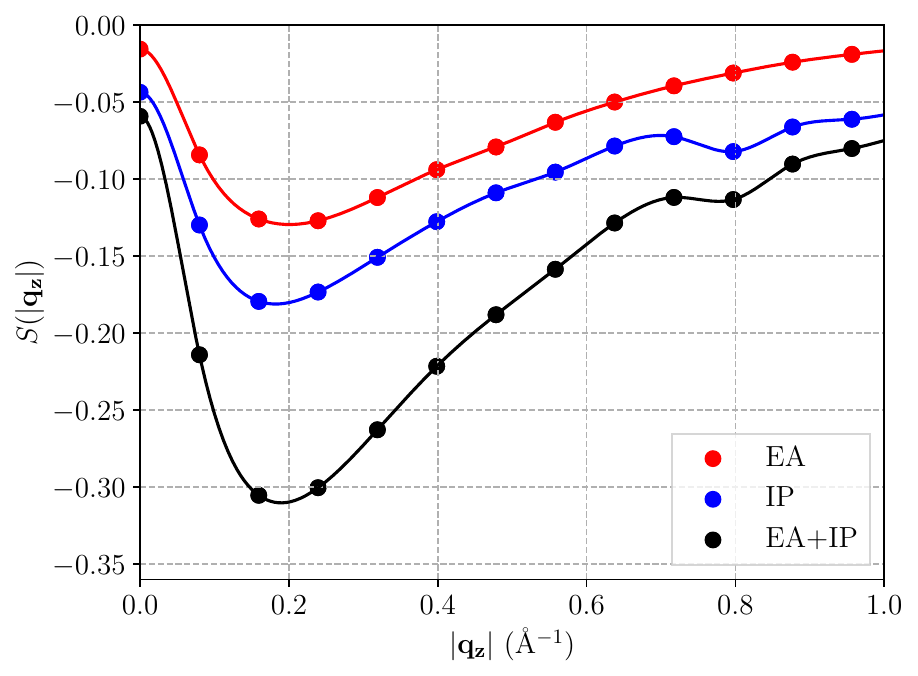}
	\caption{IP- and EA-EOM-CCSD structure factor and their sum corresponding to the band gap of tPA along the direction of the chain for a $1\times 1\times 32$ super cell ($z$-direction). As a guide to the eye, the
	calculated data points are connected via a cubic spline interpolation.}
    \label{fig:tPA-structure-factors}
\end{figure}

In order to verify the applicability
of the convergence rates, which were derived in the
long-wavelength limit ($|\mathbf{q}|\to 0$), we now investigate 
the IP- and EA-EOM-CCSD structure factors of an actual one-dimensional system, the tPA chain 
(note, that the previosuly studied LiH system, is a bulk material, however only extended in one direction). 
For that purpose, super cell sizes
of up to $1\times 1\times 32$ were computed via IP- and EA-EOM-CCSD.
The IP- and EA-EOM-CCSD structure factors, and the structure factor corresponding to the correlation
energy contribution to the band gap, that is the sum of IP
and EA in the present convention, are illustrated in Figure~\ref{fig:tPA-structure-factors}.

Figure~\ref{fig:tPA-structure-factors} shows that
for the $1\times 1\times 32$ super cell of the tPA chain, the minimum
of the EOM-CC structure factor is resolved and two data points to the
left of the minimum are obtained.
As discussed previously for the case of the LiH EOM-CC structure factor,
the analytical extrapolation expressions derived
for the asymptotic limit ($|\mathbf{q}\to 0|$) are strictly speaking only applicable for
system sizes for which the minimum of the structure factor
can be resolved. By applying a cubic spline interpolation
to the EOM-CC structure factors in Figure \ref{fig:tPA-structure-factors},
one finds that the minimum is located near  $~0.2\;\mathrm{\AA}^{-1}$, which corresponds
to approximately a $1\times 1\times 13$ \gls{bvk} cell. It is, however, important to note that the data point directly
to the right of the minimum in Figure \ref{fig:tPA-structure-factors}, which corresponds to a $1\times 1\times 10$ \gls{bvk} cell,
is in close proximity to the minimum itself, so that the quadrature error resulting from extrapolation of
system sizes as small as 10 primitive cells in one direction can be assumed to be small.
Compared to the characteristic distance of $~0.5\;\mathrm{\AA}^{-1}$ previously found for LiH,
this suggests that the length scale of the electronic correlation is more than doubled compared to LiH.
One possible explanation for that is the presence of long-ranged dispersion interactions, which are expected to be
more prominent in an organic compound like tPA than in an ionic compound with small ions like LiH.
Note, that in the case of the IP-EOM-CCSD structure factor in Figure \ref{fig:tPA-structure-factors}, we observe
a second, shallow minimum at $~0.8\;\mathrm{\AA}^{-1}$. Since this feature is located in the medium- to short
range region of the EOM-CC structure factor, this is most certainly an artefact of the comparably small PW
basis set, which had to be used to compute the $1\times 1\times 32$ super cell of tPA. As, however, we are interested
in the long-range characteristics of the electronic correlation, this is not expected to have any effect on the
properties of the tPA EOM-CCSD structure factors discussed so far.

\subsubsection{Convergence of the correlation contribution}

If the derived convergence rate in Equation
\ref{eq:1d-convergence-rate-Nk} does indeed accurately model the
convergence of one-dimensional systems to the \gls{tdl}, will be determined by
studying the convergence of the numerically determined band gap for \textit{trans}-Polyacetylene (tPA).

\begin{figure}[h]
\includegraphics[scale=0.6]{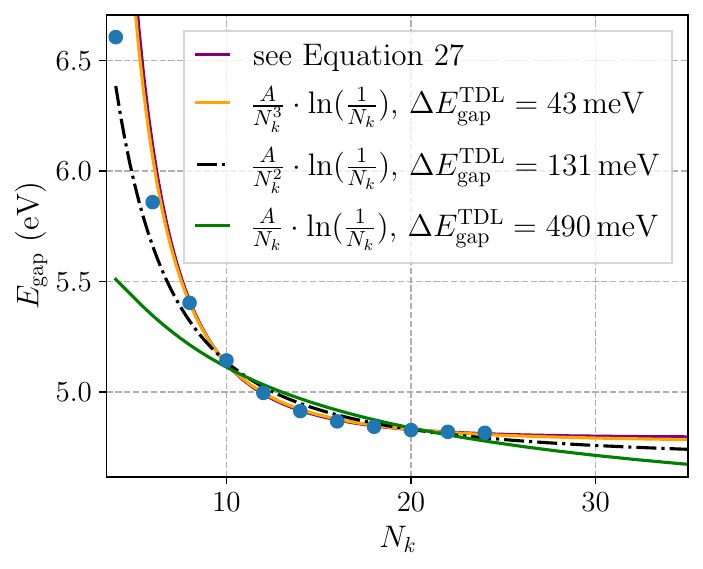}
\caption{%
  \label{fig:eom-cc-tpa-convergence}
  Application of the derived convergence model and its subsets to the
  convergence of a tPA chain's energy band gap. The models were fitted to data points with $N_k\geq 8$,
  where \( N_k \) denotes the total number of \( k \)-points used in the calculation. For the reduced models
	the deviation of the extrapolated band gap $\Delta E_{\text{gap}}^{\text{TDL}}$ from the full model (purple) is shown.}
\end{figure}

In the pursuit to determine if some order of $q$ of
$S^{\text{IP/EA}}(q)$ in Equation \ref{eq:eom-sf-asymptotic} is
dominant and if the derived convergence rate in Equation
\ref{eq:1d-convergence-rate-Nk} can be reduced meaningfully to a
single leading-order contribution, a set of convergence rate models
are applied to the finite-size convergence of tPA in Figure
\ref{fig:eom-cc-tpa-convergence}.  In particular, the full derived
model in Equation \ref{eq:1d-convergence-rate-Nk} is fitted to the
calculated band gaps as a function of $N_k$ along the direction of the
tPA chain. This model is compared to three other convergence rates,
which are the leading-order contributions originating from the $q^0$,
$q^1$ and $q^2$ contribution to the long-wavelength limit of the EOM-CC
structure factor, which are $AN_k^{-1}\ln(N_k^{-1})$,
$AN_k^{-2}\ln(N_k^{-1})$ and $AN_k^{-3}\ln(N_k^{-1})$,
respectively, where $A$ is determined by fitting to the numerical data. 
To prioritize the accurate description of the finite-size
convergence in the long-range limit, a modified least-square
cost function was used for fitting, where a data point corresponding
to a system size of $N_k$ was weighted with
a factor of $N_k^2$.
In agreement with the computed
EOM-CC structure factor of tPA shown in Figure \ref{fig:tPA-structure-factors},
it was found that the derived convergence rates only describe
the numerical data in Figure \ref{fig:eom-cc-tpa-convergence}
qualitatively well for system sizes that exceed some minimal number
of $\mathbf{k}$-points $N_k$, so that the models in 
Figure \ref{fig:eom-cc-tpa-convergence} are fitted to the
EOM-CCSD band gaps for $N_k\geq 8$. To remain consistent, all
following fits to FHI-aims band gaps, that is 
Figure \ref{fig:downsampled-eom-cc-tpa-convergence},
\ref{fig:gw-vs-eom-fs-convergence} and
\ref{fig:gw-vs-eom-fs-convergence-downsampled}, are shown for 
$N_k\geq 4$, but only fitted to data points with $N_k\geq 8$.

Undeniably, the $q^0$-order contribution to Equation
\ref{eq:1d-convergence-rate-Nk} on its own, given by \( A N_k^{-1}\ln{}N_k^{-1} \),
fails entirely to model
the computed data in Figure \ref{fig:eom-cc-tpa-convergence},
underestimating the band gap in the \gls{tdl} by almost  $500\;\text{meV}$
relative to the full model. This lends credence to the aforementioned
estimation that for quasi-particle excitations with a minor many-body
character, these $q^0$ contributions would only play a minor role in
the description of the convergence to the \gls{tdl} and is also consistent
with the small contribution of the $|\mathbf{q}|^0$ term of the
EOM-CC structure factor to the correlation energy in the
anisotropic LiH cell in Section \ref{sec:lih}. As a matter of fact,
the many-body contribution of the IP and the EA in tPA were found
to be both about $5\%$. Moreover, we stress that even if this contribution was
large, the employed Coulomb singularity treatment in FHI-aims would capture
this contribution to the finite-size error.
Similarly, the model corresponding to a EOM-CC structure factor linear
in $\mathbf{q}$, given by $AN_k^{-2}\ln(N_k^{-1})$,
converges visibly slower than the calculated data points
leading to an underestimate of the band gap of over $130\;\text{meV}$
compared to the full model.
This, in combination with the observation that the LiH EOM-CC structure factor
does not seem to exhibit any linear behavior for $q\to 0$, hints at the
possibility that in some systems -- or possibly in general -- there is no
or only a negligible linear contribution to $S^{\text{IP/EA}}(\mathbf{q})$ for $q\to 0$.
Instead, at least in the case of tPA, the $q^2$ contribution to the EOM-CC
structure factor appears to dictate the \gls{tdl} convergence of the band gap.
Even though the $AN_k^{-3}\ln(N^{-1})$ model notably deviates from the calculated
data for smaller $N_k$, it matches perfectly for $N_k=8$ to $N_k=24$,
lying virtually on top of the full model and yielding a band gap that
is underestimated by $43\;\text{meV}$.

This range of applicability is
slightly larger than the prior investigation of the EOM-CC structure factors
for tPA would suggest, were a minimum of 10 primitive cells or $\mathbf{k}$-points
in one direction were found to be necessary. One likely explanation for that
minor disagreement is the utilization of different Coulomb potential approximations
in both codes as detailed in Section \ref{sec:computational-details}.

To conclusively verify the precision and accuracy of both the EOM-CCSD band gap data obtained via FHI-aims 
and the herein proposed extrapolation approach, 
the EOM-CCSD band gap calculations for supercells of size up to $1\times 1\times 16$ were repeated using
one of the structures investigated by Windom et al.\cite{windom2022examining}. In that study, the fundamental
band gap of different tPA geometries was investigated on the
EOM-CCSD level of theory employing increasingly long
tPA oligomers with a cc-pVTZ basis set. 
For the comparison, the central C\textsubscript{2}H\textsubscript{2} unit of one of these structures, 
in the original paper denoted by tPA3,
was extracted and treated under periodic boundary conditions
in the same manner as the B3LYP-optimized structure studied
so far.
To ensure a fair comparison, the \texttt{FHI-aims} calculations
were performed employing the loc-NAO-VCC-3Z basis, the results
of which are shown in Figure \ref{fig:eom-cc-tpa-convergence}.

\begin{figure}[h]
\includegraphics[scale=0.6]{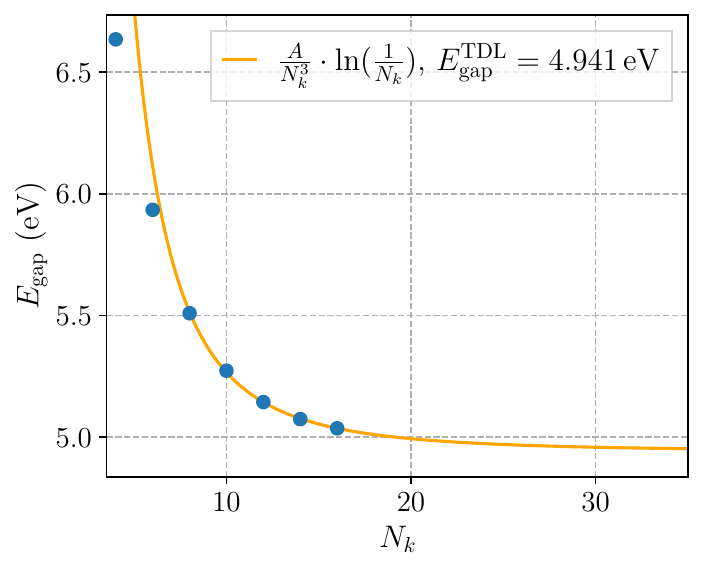}
\caption{%
  \label{fig:eom-cc-tpa-convergence}
  Application of the $AN_k^{-3}\ln(N_k^{-1})$ model to the convergence of the EOM-CCSD band gaps 
	for the tPA3 chain. The model was fitted to data with $N_k\geq 8$.}
\end{figure}

Using the previously identified $AN_k^{-3}\ln(N_k^{-1})$
leading-order model, an EOM-CCSD band gap of $4.914\,\text{eV}$
in the bulk-limit is found.
This is in reasonable agreement
with the estimated \acrshort{tdl} value of $5.07\,\text{eV}$ of the original study. The
remaining deviation of roughly $160\,\text{meV}$ is likely the
result of comparably small oligomer sizes, namely 6 to 9 C\textsubscript{2}H\textsubscript{2} units 
that were used for the extrapolation. Also, the fact that
neither of the results are
converged with respect to the basis set (see Table \ref{tab:aims-basis-set-convergence}) makes an exact
comparison more difficult.


\subsubsection{Convergence of the mean-field contributions}

\begin{figure}
    \centering
    \includegraphics[scale=0.55]{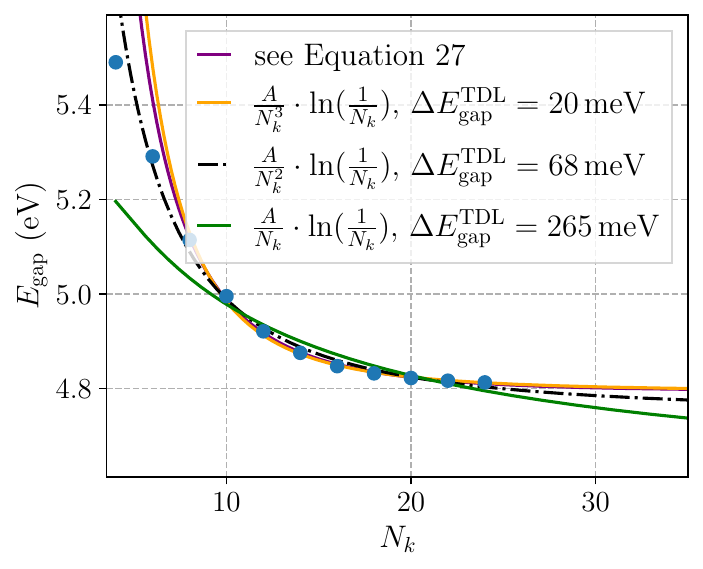}
    \caption{Application of the derived convergence model and its subsets to the convergence of the down-sampled EOM-CCSD band gaps for a tPA chain.The models were fitted to data with $N_k\geq 8$. For the reduced models
        the deviation of the extrapolated band gap $\Delta E_{\text{gap}}^{\text{TDL}}$ from the full model (purple) is shown.}
    \label{fig:downsampled-eom-cc-tpa-convergence}
\end{figure}

What has been neglected so far, however, is the potential influence of the single-body contributions
from the underlying \gls{hf} calculations on the overall
convergence of the band gap. Most importantly,
the convergence of the \gls{hf} exchange with
respect to system size needs to be considered.
So far, the \gls{hf} calculations have been performed with the same system sizes as were subsequently
used for the CC calculations, so that the total finite-size error of the final EOM-CCSD band gap contained
contributions from both the mean-field calculation and the correlated CC calculation.
To allow for a more systematic study, which only includes long-range correlation effects, the EOM-CCSD band gap results will be recomputed using converged \gls{hf} single-particle states and eigenenergies from supercells of size $1\times 1\times 48$ or more. Via this down-sampling approach, one can independently analyze the
convergence of the correlation energy of the EOM-CCSD band gaps. The finite-size convergence including the leading-order fits as they have been
shown in Figure~\ref{fig:eom-cc-tpa-convergence} is presented in Figure~\ref{fig:downsampled-eom-cc-tpa-convergence}
for the down-sampled data.

The fitted leading-order models in Figure \ref{fig:downsampled-eom-cc-tpa-convergence} suggest
that after the single-body contributions, particularly the \gls{hf} exchange, has been accounted for, still the $AN_k^{-3}\ln(N_k^{-1})$ model which results from the contribution to the structure
factor quadratic in $q$, describes the convergence
to the thermodynamic limit best and virtually lies on top of the data points for $N_k > 10$.

\subsection{Comparison to $G_0W_0$}
To test the validity of the derived convergence rate for one-dimensional systems
outside of CC theory, the leading-order contribution determined in the previous
section is applied to the $GW$ approximation, as well.
For that purpose, for the same system sizes that were previously computed via
IP- and EA-EOM-CCSD, the band gaps were obtained using the $G_0W_0$ method
with a \gls{hf} starting point ($G_0W_0$@HF) to allow for a direct comparison to
the EOM-CC results.

\begin{figure}
    \centering
    \includegraphics[width=0.5\textwidth]{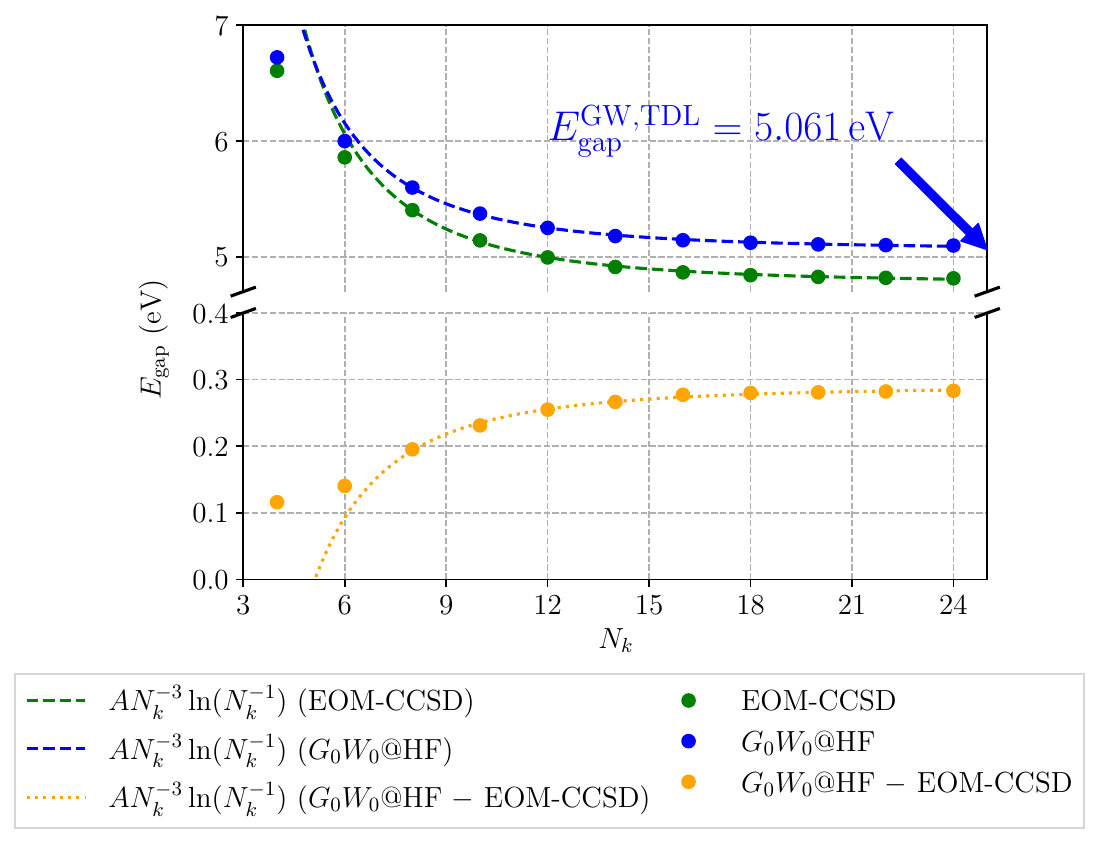}
    \caption{Comparison of the convergence to the \gls{tdl} for IP/EA-EOM-CCSD
    theory and the $G_0W_0$@HF method for a single chain of tPA. The
    band gaps of both methods are modeled via a $AN_k^{-3}\ln(N_k^{-1})$ term,
    shown to dictate the convergence to the \gls{tdl} for EOM-CC in Section \ref{sec:tPA}.
    The models were fitted to data with $N_k\geq 8$.
    }
    \label{fig:gw-vs-eom-fs-convergence}
\end{figure}

The performance of the previously extracted $AN_k^{-3}\ln(N_k^{-1})$
convergence rate applied to both EOM-CCSD and $G_0W_0$@HF is shown in
Figure \ref{fig:gw-vs-eom-fs-convergence}. There, one observes that
the convergence behavior of the $GW$ method with respect to the number
of $\mathbf{k}$-points $N_k$ almost mirrors the one of the EOM-CCSD
method. In the same way the leading-order term originating from the
$q^2$ contribution to the EOM-CC structure factor models the band gap
convergence of both theories accurately over the entire range of
$N_k=8$ to $N_k=24$.

\begin{figure}
    \centering
    \includegraphics[width=0.5\textwidth]{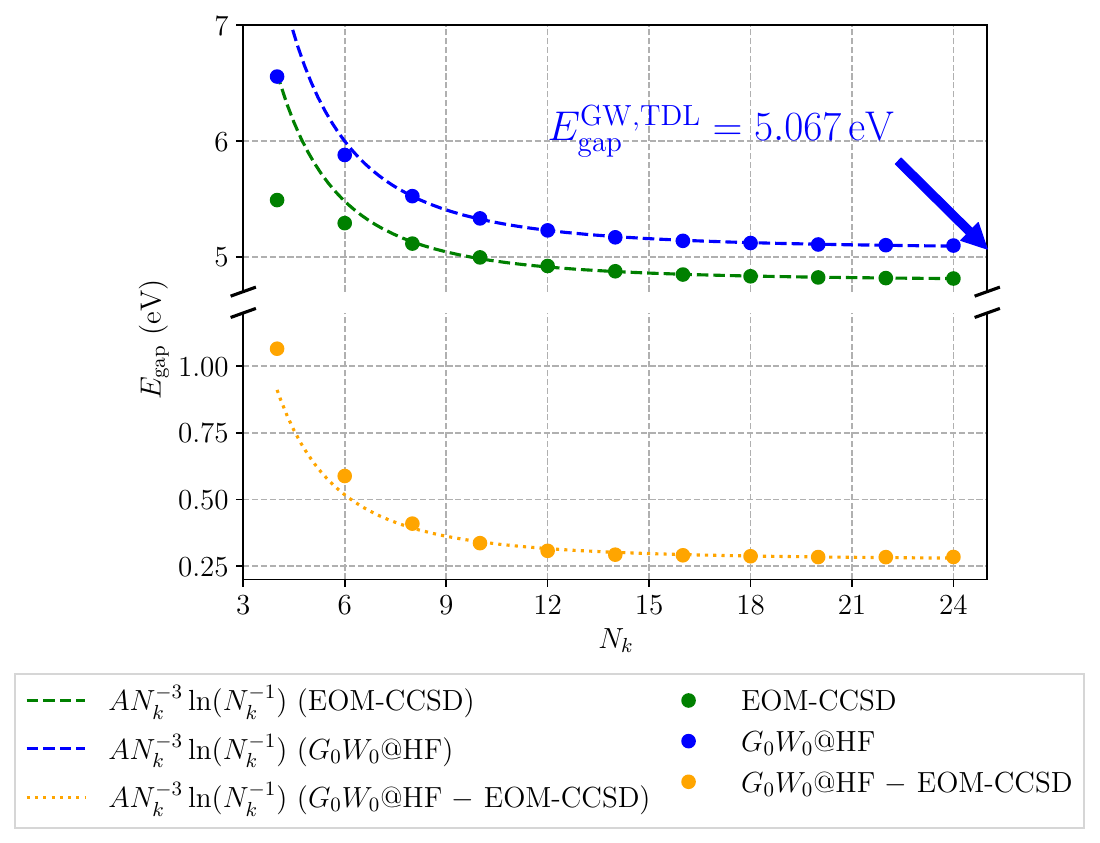}
    \caption{Comparison of the convergence to the \gls{tdl} for IP/EA-EOM-CCSD
    theory and the $G_0W_0$@HF method for a single chain of tPA
    after applying down-sampling. The
    band gaps of both methods are modeled via a $AN_k^{-3}\ln(N_k^{-1})$ term,
    shown to dictate the convergence to the \gls{tdl} for EOM-CC in Section \ref{sec:tPA}.
    The models were fitted to data with $N_k\geq 8$.
    }
    \label{fig:gw-vs-eom-fs-convergence-downsampled}
\end{figure}

In an analogous step as in Section \ref{sec:tPA}, the comparison
between EOM-CCSD and $G_0W_0$ band gaps was repeated after converging
the underlying \gls{hf} orbitals and eigenenergies via down-sampling. As previously, for
that purpose the \gls{hf} calculation was performed on a $\mathbf{k}$-grid
of size $1\times 1\times 48$ or more and the resulting \gls{hf} orbitals and eigenergies
were subsequently used to compute the IP/EA-EOM-CCSD and $G_0W_0$
band gaps on $\mathbf{k}$-meshes between $1\times 1\times 8$ and $1\times 1\times 24$.
The resulting convergence to the bulk-limit for the two methods
is shown in Figure \ref{fig:gw-vs-eom-fs-convergence-downsampled}.

Figure \ref{fig:gw-vs-eom-fs-convergence-downsampled} shows
that after the convergence of the single-body contributions of
the underlying \gls{hf} method has been ensured, the remaining
contributions to both the EOM-CCSD and $G_0W_0$ quasi-particle energies
converge with the previously identified leading-order model
$AN_k^{-3}\ln(N_k^{-1})$. While the down-sampling reduces
the individual band gaps of both methods most notably for small values
of $N_k$, the convergence behavior and the band gap in the
\gls{tdl} remain unaffected, while the extrapolated $G_0W_0$@HF
band gap in the \gls{tdl} changes by less than $6\,\text{meV}$

Figures
\ref{fig:gw-vs-eom-fs-convergence} and
\ref{fig:gw-vs-eom-fs-convergence-downsampled}
also show the differences between 
$G_0W_0$ and EOM-CCSD band gaps retrieved as a function of $N_k$ without and with applying
the down-sampling technique, respectively.
We note that the convergence of the difference can also be well
approximated using an $AN_k^{-3}\ln(N_k^{-1})$ model.
The relatively small fitting parameter indicates
that both methods capture contributions to the
correlation energy with a similar magnitude and the same leading-order behaviour.
In other words, although the $G_0W_0$ and EOM-CCSD band gaps exhibit a
finite size error that is similar in magnitude and converges with the same analytical
behavior to the \gls{tdl}, there exist
different leading-order diagrammatic contributions of both methods.
In practice one can still benefit from the similarity of both methods
by correcting the finite size error of EOM-CCSD band gaps using
the computationally significantly cheaper $G_0W_0$ method.
As noted before, the long-range behavior of
the EOM-CC structure factor is independent of the dimension of the
system. Therefore, one can infer that a $GW$-aided finite-size
correction technique for EOM-CC theory can in principle be performed
for one-, two- and three-dimensional systems alike.

An important question remaining is: What diagrammatic contributions
account for the differences between $G_0W_0$ and EOM-CCSD in the long-wavelength limit?
Although, the present work does not provide an explicit answer,
we note that previous studies performed detailed investigations of the 
relationship between the $GW$ approximation and the
EOM-CC framework.
In the first study of this sort, performed by 
Lange and Berkelbach\cite{lange2018relation}, it was found that
the $G_0W_0$@HF approximation features more higher-order ring
diagrams than the Green's function of IP- and EA-EOM-CCSD.
These diagrams are known to be particularly relevant
for the description of electronic correlation in the long-wavelength limit of the ground state
for metallic systems~\cite{masios2023, mattuck}.
Since then, different flavors of the $GW$ approximation have been
reformulated in a CC-like fashion. Bintrim and Berkelbach \cite{bintrim2021full}
presented the EOM-CC-like working equations for the $G_0W_0$ method
in the Tamm-Dancoff approximation ($G_0W_0$-TDA) resulting in a frequency-independent
method.
More recently, the exact equivalence between the $G_0W_0$@HF method and the unitary
IP/EA-EOM-CCSD method using the quasi-boson approximation has been derived~\cite{tolle2023exact}.
We also note that different Green's function methods have been formulated based 
on the CC formalism~\cite{nooijen1992coupled, backhouse2022constructing},
offering an alternative avenue to compare CC and $GW$ methods.
The above relationships shall be further analysed in future work to develop computationally efficient and
accurate finite-size corrections for EOM-CCSD band gaps.

\section{Conclusion}\label{sec:conclusion}
We have investigated the convergence of the EOM-CC band gap to the \gls{tdl}
by a formal analysis of the correlation structure factor of the IP- and
EA-EOM-CCSD method $S^{\text{IP/EA}}(\mathbf{q})$ in the long-wavelength limit
($|\mathbf{q}|\to 0$). As a result, we derived the convergence rate for
the one-, two- and three-dimensional case.  In order to verify the validity of
that approach, we focused on the one-dimensional case to be able to
compare to numerical results, using a chain of LiH unit cells and a
chain of \textit{trans}-Polyacetylene as examples. Visualizing the
EOM-CC structure factor and modeling the finite-size
convergence using the derived convergence rate for one-dimensional systems suggests
that the band gap converges to the \gls{tdl} with $AN_k^{-3}\ln(N_k^{-1})$
in leading order for $N_k \rightarrow \infty$. 
In analogy to the one-dimensional case, we expect a $AN_k^{-\frac{3}{2}}$ 
behavior for two-dimensional systems and a $AN_k^{-1}$ behavior 
for three-dimensional ones.
Finally, we verified that our findings extend
beyond EOM-CC theory and apply to the $G_0W_0$ method as well.
We find that band gaps converge with the same rate in both
theories, 
providing a formal justification to use the $GW$ approach for the extrapolation
of single-$k$-point EOM-CC results to the \gls{tdl}.

\begin{acknowledgments}
  E.M. is thankful to Min-Ye Zhang and Sebastian Kokott for valuable discussions. 
  This project was supported by TEC1p [the European Research Council (ERC) Horizon 2020 research 
  and innovation program, Grant Agreement No.740233].  
\end{acknowledgments}

\nocite{*}

\bibliography{main}

\providecommand{\noopsort}[1]{}\providecommand{\singleletter}[1]{#1}%
\begin{thebibliography}{55}%
\makeatletter
\providecommand \@ifxundefined [1]{%
 \@ifx{#1\undefined}
}%
\providecommand \@ifnum [1]{%
 \ifnum #1\expandafter \@firstoftwo
 \else \expandafter \@secondoftwo
 \fi
}%
\providecommand \@ifx [1]{%
 \ifx #1\expandafter \@firstoftwo
 \else \expandafter \@secondoftwo
 \fi
}%
\providecommand \natexlab [1]{#1}%
\providecommand \enquote  [1]{``#1''}%
\providecommand \bibnamefont  [1]{#1}%
\providecommand \bibfnamefont [1]{#1}%
\providecommand \citenamefont [1]{#1}%
\providecommand \href@noop [0]{\@secondoftwo}%
\providecommand \href [0]{\begingroup \@sanitize@url \@href}%
\providecommand \@href[1]{\@@startlink{#1}\@@href}%
\providecommand \@@href[1]{\endgroup#1\@@endlink}%
\providecommand \@sanitize@url [0]{\catcode `\\12\catcode `\$12\catcode
  `\&12\catcode `\#12\catcode `\^12\catcode `\_12\catcode `\%12\relax}%
\providecommand \@@startlink[1]{}%
\providecommand \@@endlink[0]{}%
\providecommand \url  [0]{\begingroup\@sanitize@url \@url }%
\providecommand \@url [1]{\endgroup\@href {#1}{\urlprefix }}%
\providecommand \urlprefix  [0]{URL }%
\providecommand \Eprint [0]{\href }%
\providecommand \doibase [0]{http://dx.doi.org/}%
\providecommand \selectlanguage [0]{\@gobble}%
\providecommand \bibinfo  [0]{\@secondoftwo}%
\providecommand \bibfield  [0]{\@secondoftwo}%
\providecommand \translation [1]{[#1]}%
\providecommand \BibitemOpen [0]{}%
\providecommand \bibitemStop [0]{}%
\providecommand \bibitemNoStop [0]{.\EOS\space}%
\providecommand \EOS [0]{\spacefactor3000\relax}%
\providecommand \BibitemShut  [1]{\csname bibitem#1\endcsname}%
\let\auto@bib@innerbib\@empty
\bibitem [{\citenamefont {Perdew}\ \emph {et~al.}(2017)\citenamefont {Perdew},
  \citenamefont {Yang}, \citenamefont {Burke}, \citenamefont {Yang},
  \citenamefont {Gross}, \citenamefont {Scheffler}, \citenamefont {Scuseria},
  \citenamefont {Henderson}, \citenamefont {Zhang}, \citenamefont {Ruzsinszky}
  \emph {et~al.}}]{perdew2017understanding}%
  \BibitemOpen
  \bibfield  {author} {\bibinfo {author} {\bibfnamefont {J.~P.}\ \bibnamefont
  {Perdew}}, \bibinfo {author} {\bibfnamefont {W.}~\bibnamefont {Yang}},
  \bibinfo {author} {\bibfnamefont {K.}~\bibnamefont {Burke}}, \bibinfo
  {author} {\bibfnamefont {Z.}~\bibnamefont {Yang}}, \bibinfo {author}
  {\bibfnamefont {E.~K.}\ \bibnamefont {Gross}}, \bibinfo {author}
  {\bibfnamefont {M.}~\bibnamefont {Scheffler}}, \bibinfo {author}
  {\bibfnamefont {G.~E.}\ \bibnamefont {Scuseria}}, \bibinfo {author}
  {\bibfnamefont {T.~M.}\ \bibnamefont {Henderson}}, \bibinfo {author}
  {\bibfnamefont {I.~Y.}\ \bibnamefont {Zhang}}, \bibinfo {author}
  {\bibfnamefont {A.}~\bibnamefont {Ruzsinszky}},  \emph {et~al.},\ }\bibfield
  {title} {\enquote {\bibinfo {title} {Understanding band gaps of solids in
  generalized kohn--sham theory},}\ }\href@noop {} {\bibfield  {journal}
  {\bibinfo  {journal} {Proceedings of the national academy of sciences}\
  }\textbf {\bibinfo {volume} {114}},\ \bibinfo {pages} {2801} (\bibinfo {year}
  {2017})}\BibitemShut {NoStop}%
\bibitem [{\citenamefont {Hedin}(1965)}]{hedin1965new}%
  \BibitemOpen
  \bibfield  {author} {\bibinfo {author} {\bibfnamefont {L.}~\bibnamefont
  {Hedin}},\ }\bibfield  {title} {\enquote {\bibinfo {title} {New method for
  calculating the one-particle green's function with application to the
  electron-gas problem},}\ }\href@noop {} {\bibfield  {journal} {\bibinfo
  {journal} {Physical Review}\ }\textbf {\bibinfo {volume} {139}},\ \bibinfo
  {pages} {A796} (\bibinfo {year} {1965})}\BibitemShut {NoStop}%
\bibitem [{\citenamefont {Golze}\ \emph {et~al.}(2019)\citenamefont {Golze},
  \citenamefont {Dvorak},\ and\ \citenamefont {Rinke}}]{golze2019gw}%
  \BibitemOpen
  \bibfield  {author} {\bibinfo {author} {\bibfnamefont {D.}~\bibnamefont
  {Golze}}, \bibinfo {author} {\bibfnamefont {M.}~\bibnamefont {Dvorak}}, \
  and\ \bibinfo {author} {\bibfnamefont {P.}~\bibnamefont {Rinke}},\ }\bibfield
   {title} {\enquote {\bibinfo {title} {The gw compendium: A practical guide to
  theoretical photoemission spectroscopy},}\ }\href@noop {} {\bibfield
  {journal} {\bibinfo  {journal} {Frontiers in chemistry}\ }\textbf {\bibinfo
  {volume} {7}},\ \bibinfo {pages} {377} (\bibinfo {year} {2019})}\BibitemShut
  {NoStop}%
\bibitem [{\citenamefont {Shishkin}\ \emph {et~al.}(2007)\citenamefont
  {Shishkin}, \citenamefont {Marsman},\ and\ \citenamefont
  {Kresse}}]{shishkin2007accurate}%
  \BibitemOpen
  \bibfield  {author} {\bibinfo {author} {\bibfnamefont {M.}~\bibnamefont
  {Shishkin}}, \bibinfo {author} {\bibfnamefont {M.}~\bibnamefont {Marsman}}, \
  and\ \bibinfo {author} {\bibfnamefont {G.}~\bibnamefont {Kresse}},\
  }\bibfield  {title} {\enquote {\bibinfo {title} {Accurate quasiparticle
  spectra from self-consistent gw calculations with vertex corrections},}\
  }\href@noop {} {\bibfield  {journal} {\bibinfo  {journal} {Physical review
  letters}\ }\textbf {\bibinfo {volume} {99}},\ \bibinfo {pages} {246403}
  (\bibinfo {year} {2007})}\BibitemShut {NoStop}%
\bibitem [{\citenamefont {Gr\"uneis}\ \emph {et~al.}(2014)\citenamefont
  {Gr\"uneis}, \citenamefont {Kresse}, \citenamefont {Hinuma},\ and\
  \citenamefont {Oba}}]{Gruneis2014}%
  \BibitemOpen
  \bibfield  {author} {\bibinfo {author} {\bibfnamefont {A.}~\bibnamefont
  {Gr\"uneis}}, \bibinfo {author} {\bibfnamefont {G.}~\bibnamefont {Kresse}},
  \bibinfo {author} {\bibfnamefont {Y.}~\bibnamefont {Hinuma}}, \ and\ \bibinfo
  {author} {\bibfnamefont {F.}~\bibnamefont {Oba}},\ }\bibfield  {title}
  {\enquote {\bibinfo {title} {Ionization potentials of solids: The importance
  of vertex corrections},}\ }\href {\doibase 10.1103/PhysRevLett.112.096401}
  {\bibfield  {journal} {\bibinfo  {journal} {Phys. Rev. Lett.}\ }\textbf
  {\bibinfo {volume} {112}},\ \bibinfo {pages} {096401} (\bibinfo {year}
  {2014})}\BibitemShut {NoStop}%
\bibitem [{\citenamefont {Lewis}\ and\ \citenamefont
  {Berkelbach}(2019)}]{lewis2019vertex}%
  \BibitemOpen
  \bibfield  {author} {\bibinfo {author} {\bibfnamefont {A.~M.}\ \bibnamefont
  {Lewis}}\ and\ \bibinfo {author} {\bibfnamefont {T.~C.}\ \bibnamefont
  {Berkelbach}},\ }\bibfield  {title} {\enquote {\bibinfo {title} {Vertex
  corrections to the polarizability do not improve the gw approximation for the
  ionization potential of molecules},}\ }\href@noop {} {\bibfield  {journal}
  {\bibinfo  {journal} {Journal of Chemical Theory and Computation}\ }\textbf
  {\bibinfo {volume} {15}},\ \bibinfo {pages} {2925} (\bibinfo {year}
  {2019})}\BibitemShut {NoStop}%
\bibitem [{\citenamefont {Kutepov}(2022)}]{kutepov2022full}%
  \BibitemOpen
  \bibfield  {author} {\bibinfo {author} {\bibfnamefont {A.~L.}\ \bibnamefont
  {Kutepov}},\ }\bibfield  {title} {\enquote {\bibinfo {title} {Full versus
  quasiparticle self-consistency in vertex-corrected gw approaches},}\
  }\href@noop {} {\bibfield  {journal} {\bibinfo  {journal} {Physical Review
  B}\ }\textbf {\bibinfo {volume} {105}},\ \bibinfo {pages} {045124} (\bibinfo
  {year} {2022})}\BibitemShut {NoStop}%
\bibitem [{\citenamefont {Romaniello}\ \emph {et~al.}(2012)\citenamefont
  {Romaniello}, \citenamefont {Bechstedt},\ and\ \citenamefont
  {Reining}}]{romaniello2012beyond}%
  \BibitemOpen
  \bibfield  {author} {\bibinfo {author} {\bibfnamefont {P.}~\bibnamefont
  {Romaniello}}, \bibinfo {author} {\bibfnamefont {F.}~\bibnamefont
  {Bechstedt}}, \ and\ \bibinfo {author} {\bibfnamefont {L.}~\bibnamefont
  {Reining}},\ }\bibfield  {title} {\enquote {\bibinfo {title} {Beyond the g w
  approximation: Combining correlation channels},}\ }\href@noop {} {\bibfield
  {journal} {\bibinfo  {journal} {Physical Review B}\ }\textbf {\bibinfo
  {volume} {85}},\ \bibinfo {pages} {155131} (\bibinfo {year}
  {2012})}\BibitemShut {NoStop}%
\bibitem [{\citenamefont {Stanton}\ and\ \citenamefont
  {Bartlett}(1993)}]{stanton1993equation}%
  \BibitemOpen
  \bibfield  {author} {\bibinfo {author} {\bibfnamefont {J.~F.}\ \bibnamefont
  {Stanton}}\ and\ \bibinfo {author} {\bibfnamefont {R.~J.}\ \bibnamefont
  {Bartlett}},\ }\bibfield  {title} {\enquote {\bibinfo {title} {The equation
  of motion coupled-cluster method. a systematic biorthogonal approach to
  molecular excitation energies, transition probabilities, and excited state
  properties},}\ }\href@noop {} {\bibfield  {journal} {\bibinfo  {journal} {The
  Journal of chemical physics}\ }\textbf {\bibinfo {volume} {98}},\ \bibinfo
  {pages} {7029} (\bibinfo {year} {1993})}\BibitemShut {NoStop}%
\bibitem [{\citenamefont {McClain}\ \emph {et~al.}(2017)\citenamefont
  {McClain}, \citenamefont {Sun}, \citenamefont {Chan},\ and\ \citenamefont
  {Berkelbach}}]{mcclain2017gaussian}%
  \BibitemOpen
  \bibfield  {author} {\bibinfo {author} {\bibfnamefont {J.}~\bibnamefont
  {McClain}}, \bibinfo {author} {\bibfnamefont {Q.}~\bibnamefont {Sun}},
  \bibinfo {author} {\bibfnamefont {G.~K.-L.}\ \bibnamefont {Chan}}, \ and\
  \bibinfo {author} {\bibfnamefont {T.~C.}\ \bibnamefont {Berkelbach}},\
  }\bibfield  {title} {\enquote {\bibinfo {title} {Gaussian-based
  coupled-cluster theory for the ground-state and band structure of solids},}\
  }\href@noop {} {\bibfield  {journal} {\bibinfo  {journal} {Journal of
  chemical theory and computation}\ }\textbf {\bibinfo {volume} {13}},\
  \bibinfo {pages} {1209} (\bibinfo {year} {2017})}\BibitemShut {NoStop}%
\bibitem [{\citenamefont {Gao}\ \emph {et~al.}(2020)\citenamefont {Gao},
  \citenamefont {Sun}, \citenamefont {Jason}, \citenamefont {Motta},
  \citenamefont {McClain}, \citenamefont {White}, \citenamefont {Minnich},\
  and\ \citenamefont {Chan}}]{gao2020electronic}%
  \BibitemOpen
  \bibfield  {author} {\bibinfo {author} {\bibfnamefont {Y.}~\bibnamefont
  {Gao}}, \bibinfo {author} {\bibfnamefont {Q.}~\bibnamefont {Sun}}, \bibinfo
  {author} {\bibfnamefont {M.~Y.}\ \bibnamefont {Jason}}, \bibinfo {author}
  {\bibfnamefont {M.}~\bibnamefont {Motta}}, \bibinfo {author} {\bibfnamefont
  {J.}~\bibnamefont {McClain}}, \bibinfo {author} {\bibfnamefont {A.~F.}\
  \bibnamefont {White}}, \bibinfo {author} {\bibfnamefont {A.~J.}\ \bibnamefont
  {Minnich}}, \ and\ \bibinfo {author} {\bibfnamefont {G.~K.-L.}\ \bibnamefont
  {Chan}},\ }\bibfield  {title} {\enquote {\bibinfo {title} {Electronic
  structure of bulk manganese oxide and nickel oxide from coupled cluster
  theory},}\ }\href@noop {} {\bibfield  {journal} {\bibinfo  {journal}
  {Physical Review B}\ }\textbf {\bibinfo {volume} {101}},\ \bibinfo {pages}
  {165138} (\bibinfo {year} {2020})}\BibitemShut {NoStop}%
\bibitem [{\citenamefont {Lange}\ and\ \citenamefont
  {Berkelbach}(2018)}]{lange2018relation}%
  \BibitemOpen
  \bibfield  {author} {\bibinfo {author} {\bibfnamefont {M.~F.}\ \bibnamefont
  {Lange}}\ and\ \bibinfo {author} {\bibfnamefont {T.~C.}\ \bibnamefont
  {Berkelbach}},\ }\bibfield  {title} {\enquote {\bibinfo {title} {On the
  relation between equation-of-motion coupled-cluster theory and the gw
  approximation},}\ }\href@noop {} {\bibfield  {journal} {\bibinfo  {journal}
  {Journal of chemical theory and computation}\ }\textbf {\bibinfo {volume}
  {14}},\ \bibinfo {pages} {4224} (\bibinfo {year} {2018})}\BibitemShut
  {NoStop}%
\bibitem [{\citenamefont {McClain}\ \emph {et~al.}(2016)\citenamefont
  {McClain}, \citenamefont {Lischner}, \citenamefont {Watson}, \citenamefont
  {Matthews}, \citenamefont {Ronca}, \citenamefont {Louie}, \citenamefont
  {Berkelbach},\ and\ \citenamefont {Chan}}]{mcclain2016spectral}%
  \BibitemOpen
  \bibfield  {author} {\bibinfo {author} {\bibfnamefont {J.}~\bibnamefont
  {McClain}}, \bibinfo {author} {\bibfnamefont {J.}~\bibnamefont {Lischner}},
  \bibinfo {author} {\bibfnamefont {T.}~\bibnamefont {Watson}}, \bibinfo
  {author} {\bibfnamefont {D.~A.}\ \bibnamefont {Matthews}}, \bibinfo {author}
  {\bibfnamefont {E.}~\bibnamefont {Ronca}}, \bibinfo {author} {\bibfnamefont
  {S.~G.}\ \bibnamefont {Louie}}, \bibinfo {author} {\bibfnamefont {T.~C.}\
  \bibnamefont {Berkelbach}}, \ and\ \bibinfo {author} {\bibfnamefont
  {G.~K.-L.}\ \bibnamefont {Chan}},\ }\bibfield  {title} {\enquote {\bibinfo
  {title} {Spectral functions of the uniform electron gas via coupled-cluster
  theory and comparison to the g w and related approximations},}\ }\href@noop
  {} {\bibfield  {journal} {\bibinfo  {journal} {Physical Review B}\ }\textbf
  {\bibinfo {volume} {93}},\ \bibinfo {pages} {235139} (\bibinfo {year}
  {2016})}\BibitemShut {NoStop}%
\bibitem [{\citenamefont {T{\"o}lle}\ and\ \citenamefont
  {Kin-Lic~Chan}(2023)}]{tolle2023exact}%
  \BibitemOpen
  \bibfield  {author} {\bibinfo {author} {\bibfnamefont {J.}~\bibnamefont
  {T{\"o}lle}}\ and\ \bibinfo {author} {\bibfnamefont {G.}~\bibnamefont
  {Kin-Lic~Chan}},\ }\bibfield  {title} {\enquote {\bibinfo {title} {Exact
  relationships between the gw approximation and equation-of-motion
  coupled-cluster theories through the quasi-boson formalism},}\ }\href@noop {}
  {\bibfield  {journal} {\bibinfo  {journal} {The Journal of Chemical Physics}\
  }\textbf {\bibinfo {volume} {158}} (\bibinfo {year} {2023})}\BibitemShut
  {NoStop}%
\bibitem [{\citenamefont {Hybertsen}\ and\ \citenamefont
  {Louie}(1987)}]{hybertsen1987ab}%
  \BibitemOpen
  \bibfield  {author} {\bibinfo {author} {\bibfnamefont {M.~S.}\ \bibnamefont
  {Hybertsen}}\ and\ \bibinfo {author} {\bibfnamefont {S.~G.}\ \bibnamefont
  {Louie}},\ }\bibfield  {title} {\enquote {\bibinfo {title} {Ab initio static
  dielectric matrices from the density-functional approach. i. formulation and
  application to semiconductors and insulators},}\ }\href@noop {} {\bibfield
  {journal} {\bibinfo  {journal} {Physical Review B}\ }\textbf {\bibinfo
  {volume} {35}},\ \bibinfo {pages} {5585} (\bibinfo {year}
  {1987})}\BibitemShut {NoStop}%
\bibitem [{\citenamefont {Hunt}\ \emph {et~al.}(2018)\citenamefont {Hunt},
  \citenamefont {Szyniszewski}, \citenamefont {Prayogo}, \citenamefont
  {Maezono},\ and\ \citenamefont {Drummond}}]{hunt2018quantum}%
  \BibitemOpen
  \bibfield  {author} {\bibinfo {author} {\bibfnamefont {R.~J.}\ \bibnamefont
  {Hunt}}, \bibinfo {author} {\bibfnamefont {M.}~\bibnamefont {Szyniszewski}},
  \bibinfo {author} {\bibfnamefont {G.~I.}\ \bibnamefont {Prayogo}}, \bibinfo
  {author} {\bibfnamefont {R.}~\bibnamefont {Maezono}}, \ and\ \bibinfo
  {author} {\bibfnamefont {N.~D.}\ \bibnamefont {Drummond}},\ }\bibfield
  {title} {\enquote {\bibinfo {title} {Quantum monte carlo calculations of
  energy gaps from first principles},}\ }\href@noop {} {\bibfield  {journal}
  {\bibinfo  {journal} {Physical Review B}\ }\textbf {\bibinfo {volume} {98}},\
  \bibinfo {pages} {075122} (\bibinfo {year} {2018})}\BibitemShut {NoStop}%
\bibitem [{\citenamefont {Hunt}\ \emph {et~al.}(2020)\citenamefont {Hunt},
  \citenamefont {Monserrat}, \citenamefont {Z{\'o}lyomi},\ and\ \citenamefont
  {Drummond}}]{hunt2020diffusion}%
  \BibitemOpen
  \bibfield  {author} {\bibinfo {author} {\bibfnamefont {R.~J.}\ \bibnamefont
  {Hunt}}, \bibinfo {author} {\bibfnamefont {B.}~\bibnamefont {Monserrat}},
  \bibinfo {author} {\bibfnamefont {V.}~\bibnamefont {Z{\'o}lyomi}}, \ and\
  \bibinfo {author} {\bibfnamefont {N.}~\bibnamefont {Drummond}},\ }\bibfield
  {title} {\enquote {\bibinfo {title} {Diffusion quantum monte carlo and g w
  study of the electronic properties of monolayer and bulk hexagonal boron
  nitride},}\ }\href@noop {} {\bibfield  {journal} {\bibinfo  {journal}
  {Physical Review B}\ }\textbf {\bibinfo {volume} {101}},\ \bibinfo {pages}
  {205115} (\bibinfo {year} {2020})}\BibitemShut {NoStop}%
\bibitem [{\citenamefont {Sekino}\ and\ \citenamefont
  {Bartlett}(1984)}]{sekino1984linear}%
  \BibitemOpen
  \bibfield  {author} {\bibinfo {author} {\bibfnamefont {H.}~\bibnamefont
  {Sekino}}\ and\ \bibinfo {author} {\bibfnamefont {R.~J.}\ \bibnamefont
  {Bartlett}},\ }\bibfield  {title} {\enquote {\bibinfo {title} {A linear
  response, coupled-cluster theory for excitation energy},}\ }\href@noop {}
  {\bibfield  {journal} {\bibinfo  {journal} {International Journal of Quantum
  Chemistry}\ }\textbf {\bibinfo {volume} {26}},\ \bibinfo {pages} {255}
  (\bibinfo {year} {1984})}\BibitemShut {NoStop}%
\bibitem [{\citenamefont {Liao}\ and\ \citenamefont
  {Gr{\"u}neis}(2016)}]{liao2016communication}%
  \BibitemOpen
  \bibfield  {author} {\bibinfo {author} {\bibfnamefont {K.}~\bibnamefont
  {Liao}}\ and\ \bibinfo {author} {\bibfnamefont {A.}~\bibnamefont
  {Gr{\"u}neis}},\ }\bibfield  {title} {\enquote {\bibinfo {title}
  {Communication: Finite size correction in periodic coupled cluster theory
  calculations of solids},}\ }\href@noop {} {\bibfield  {journal} {\bibinfo
  {journal} {The Journal of Chemical Physics}\ }\textbf {\bibinfo {volume}
  {145}} (\bibinfo {year} {2016})}\BibitemShut {NoStop}%
\bibitem [{\citenamefont {Gruber}\ \emph {et~al.}(2018)\citenamefont {Gruber},
  \citenamefont {Liao}, \citenamefont {Tsatsoulis}, \citenamefont {Hummel},\
  and\ \citenamefont {Gr{\"u}neis}}]{gruber2018applying}%
  \BibitemOpen
  \bibfield  {author} {\bibinfo {author} {\bibfnamefont {T.}~\bibnamefont
  {Gruber}}, \bibinfo {author} {\bibfnamefont {K.}~\bibnamefont {Liao}},
  \bibinfo {author} {\bibfnamefont {T.}~\bibnamefont {Tsatsoulis}}, \bibinfo
  {author} {\bibfnamefont {F.}~\bibnamefont {Hummel}}, \ and\ \bibinfo {author}
  {\bibfnamefont {A.}~\bibnamefont {Gr{\"u}neis}},\ }\bibfield  {title}
  {\enquote {\bibinfo {title} {Applying the coupled-cluster ansatz to solids
  and surfaces in the thermodynamic limit},}\ }\href@noop {} {\bibfield
  {journal} {\bibinfo  {journal} {Physical Review X}\ }\textbf {\bibinfo
  {volume} {8}},\ \bibinfo {pages} {021043} (\bibinfo {year}
  {2018})}\BibitemShut {NoStop}%
\bibitem [{\citenamefont {Dittmer}\ \emph {et~al.}(2019)\citenamefont
  {Dittmer}, \citenamefont {Izsak}, \citenamefont {Neese},\ and\ \citenamefont
  {Maganas}}]{dittmer2019accurate}%
  \BibitemOpen
  \bibfield  {author} {\bibinfo {author} {\bibfnamefont {A.}~\bibnamefont
  {Dittmer}}, \bibinfo {author} {\bibfnamefont {R.}~\bibnamefont {Izsak}},
  \bibinfo {author} {\bibfnamefont {F.}~\bibnamefont {Neese}}, \ and\ \bibinfo
  {author} {\bibfnamefont {D.}~\bibnamefont {Maganas}},\ }\bibfield  {title}
  {\enquote {\bibinfo {title} {Accurate band gap predictions of semiconductors
  in the framework of the similarity transformed equation of motion coupled
  cluster theory},}\ }\href@noop {} {\bibfield  {journal} {\bibinfo  {journal}
  {Inorganic chemistry}\ }\textbf {\bibinfo {volume} {58}},\ \bibinfo {pages}
  {9303} (\bibinfo {year} {2019})}\BibitemShut {NoStop}%
\bibitem [{\citenamefont {Gallo}\ \emph {et~al.}(2021)\citenamefont {Gallo},
  \citenamefont {Hummel}, \citenamefont {Irmler},\ and\ \citenamefont
  {Gr{\"u}neis}}]{Gallo2021}%
  \BibitemOpen
  \bibfield  {author} {\bibinfo {author} {\bibfnamefont {A.}~\bibnamefont
  {Gallo}}, \bibinfo {author} {\bibfnamefont {F.}~\bibnamefont {Hummel}},
  \bibinfo {author} {\bibfnamefont {A.}~\bibnamefont {Irmler}}, \ and\ \bibinfo
  {author} {\bibfnamefont {A.}~\bibnamefont {Gr{\"u}neis}},\ }\bibfield
  {title} {\enquote {\bibinfo {title} {A periodic equation-of-motion
  coupled-cluster implementation applied to f-centers in alkaline earth
  oxides},}\ }\href {\doibase 10.1063/5.0035425} {\bibfield  {journal}
  {\bibinfo  {journal} {J. Chem. Phys.}\ }\textbf {\bibinfo {volume} {154}},\
  \bibinfo {pages} {064106} (\bibinfo {year} {2021})}\BibitemShut {NoStop}%
\bibitem [{\citenamefont {Vo}\ \emph {et~al.}(2024)\citenamefont {Vo},
  \citenamefont {Wang},\ and\ \citenamefont {Berkelbach}}]{Vo2024}%
  \BibitemOpen
  \bibfield  {author} {\bibinfo {author} {\bibfnamefont {E.~A.}\ \bibnamefont
  {Vo}}, \bibinfo {author} {\bibfnamefont {X.}~\bibnamefont {Wang}}, \ and\
  \bibinfo {author} {\bibfnamefont {T.~C.}\ \bibnamefont {Berkelbach}},\
  }\bibfield  {title} {\enquote {\bibinfo {title} {{Performance of periodic
  EOM-CCSD for bandgaps of inorganic semiconductors and insulators}},}\ }\href
  {\doibase 10.1063/5.0187856} {\bibfield  {journal} {\bibinfo  {journal} {The
  Journal of Chemical Physics}\ }\textbf {\bibinfo {volume} {160}},\ \bibinfo
  {pages} {044106} (\bibinfo {year} {2024})}\BibitemShut {NoStop}%
\bibitem [{\citenamefont {Xing}\ and\ \citenamefont
  {Lin}(2024)}]{xing2024inverse}%
  \BibitemOpen
  \bibfield  {author} {\bibinfo {author} {\bibfnamefont {X.}~\bibnamefont
  {Xing}}\ and\ \bibinfo {author} {\bibfnamefont {L.}~\bibnamefont {Lin}},\
  }\bibfield  {title} {\enquote {\bibinfo {title} {Inverse volume scaling of
  finite-size error in periodic coupled cluster theory},}\ }\href@noop {}
  {\bibfield  {journal} {\bibinfo  {journal} {Physical Review X}\ }\textbf
  {\bibinfo {volume} {14}},\ \bibinfo {pages} {011059} (\bibinfo {year}
  {2024})}\BibitemShut {NoStop}%
\bibitem [{\citenamefont {Hirao}\ and\ \citenamefont
  {Nakatsuji}(1982)}]{Hirao1982}%
  \BibitemOpen
  \bibfield  {author} {\bibinfo {author} {\bibfnamefont {K.}~\bibnamefont
  {Hirao}}\ and\ \bibinfo {author} {\bibfnamefont {H.}~\bibnamefont
  {Nakatsuji}},\ }\bibfield  {title} {\enquote {\bibinfo {title} {A
  generalization of the davidson's method to large nonsymmetric eigenvalue
  problems},}\ }\href {\doibase 10.1016/0021-9991(82)90119-X} {\bibfield
  {journal} {\bibinfo  {journal} {Journal of Computational Physics}\ }\textbf
  {\bibinfo {volume} {45}},\ \bibinfo {pages} {246} (\bibinfo {year}
  {1982})}\BibitemShut {NoStop}%
\bibitem [{\citenamefont {Mihaila}(2011)}]{mihaila2011lindhard}%
  \BibitemOpen
  \bibfield  {author} {\bibinfo {author} {\bibfnamefont {B.}~\bibnamefont
  {Mihaila}},\ }\bibfield  {title} {\enquote {\bibinfo {title} {Lindhard
  function of a d-dimensional fermi gas},}\ }\href@noop {} {\bibfield
  {journal} {\bibinfo  {journal} {arXiv preprint arXiv:1111.5337}\ } (\bibinfo
  {year} {2011})}\BibitemShut {NoStop}%
\bibitem [{\citenamefont {Sundararaman}\ and\ \citenamefont
  {Arias}(2013)}]{Sundararaman2013}%
  \BibitemOpen
  \bibfield  {author} {\bibinfo {author} {\bibfnamefont {R.}~\bibnamefont
  {Sundararaman}}\ and\ \bibinfo {author} {\bibfnamefont {T.~A.}\ \bibnamefont
  {Arias}},\ }\bibfield  {title} {\enquote {\bibinfo {title} {Regularization of
  the coulomb singularity in exact exchange by wigner-seitz truncated
  interactions: Towards chemical accuracy in nontrivial systems},}\ }\href
  {\doibase 10.1103/PhysRevB.87.165122} {\bibfield  {journal} {\bibinfo
  {journal} {Phys. Rev. B}\ }\textbf {\bibinfo {volume} {87}},\ \bibinfo
  {pages} {165122} (\bibinfo {year} {2013})}\BibitemShut {NoStop}%
\bibitem [{\citenamefont {Spencer}\ and\ \citenamefont
  {Alavi}(2008)}]{Spencer2008}%
  \BibitemOpen
  \bibfield  {author} {\bibinfo {author} {\bibfnamefont {J.}~\bibnamefont
  {Spencer}}\ and\ \bibinfo {author} {\bibfnamefont {A.}~\bibnamefont
  {Alavi}},\ }\bibfield  {title} {\enquote {\bibinfo {title} {Efficient
  calculation of the exact exchange energy in periodic systems using a
  truncated coulomb potential},}\ }\href {\doibase 10.1103/PhysRevB.77.193110}
  {\bibfield  {journal} {\bibinfo  {journal} {Phys. Rev. B}\ }\textbf {\bibinfo
  {volume} {77}},\ \bibinfo {pages} {193110} (\bibinfo {year}
  {2008})}\BibitemShut {NoStop}%
\bibitem [{\citenamefont {Carrier}\ \emph {et~al.}(2007)\citenamefont
  {Carrier}, \citenamefont {Rohra},\ and\ \citenamefont
  {G\"orling}}]{Carrier2007}%
  \BibitemOpen
  \bibfield  {author} {\bibinfo {author} {\bibfnamefont {P.}~\bibnamefont
  {Carrier}}, \bibinfo {author} {\bibfnamefont {S.}~\bibnamefont {Rohra}}, \
  and\ \bibinfo {author} {\bibfnamefont {A.}~\bibnamefont {G\"orling}},\
  }\bibfield  {title} {\enquote {\bibinfo {title} {General treatment of the
  singularities in hartree-fock and exact-exchange kohn-sham methods for
  solids},}\ }\href {\doibase 10.1103/PhysRevB.75.205126} {\bibfield  {journal}
  {\bibinfo  {journal} {Phys. Rev. B}\ }\textbf {\bibinfo {volume} {75}},\
  \bibinfo {pages} {205126} (\bibinfo {year} {2007})}\BibitemShut {NoStop}%
\bibitem [{\citenamefont {Gygi}\ and\ \citenamefont
  {Baldereschi}(1986)}]{Gygi1986}%
  \BibitemOpen
  \bibfield  {author} {\bibinfo {author} {\bibfnamefont {F.}~\bibnamefont
  {Gygi}}\ and\ \bibinfo {author} {\bibfnamefont {A.}~\bibnamefont
  {Baldereschi}},\ }\bibfield  {title} {\enquote {\bibinfo {title}
  {Self-consistent hartree-fock and screened-exchange calculations in solids:
  Application to silicon},}\ }\href {\doibase 10.1103/PhysRevB.34.4405}
  {\bibfield  {journal} {\bibinfo  {journal} {Phys. Rev. B}\ }\textbf {\bibinfo
  {volume} {34}},\ \bibinfo {pages} {4405} (\bibinfo {year}
  {1986})}\BibitemShut {NoStop}%
\bibitem [{\citenamefont {Sch{\"a}fer}\ \emph {et~al.}(2024)\citenamefont
  {Sch{\"a}fer}, \citenamefont {Van~Benschoten}, \citenamefont {Shepherd},\
  and\ \citenamefont {Gr{\"u}neis}}]{schafer2024sampling}%
  \BibitemOpen
  \bibfield  {author} {\bibinfo {author} {\bibfnamefont {T.}~\bibnamefont
  {Sch{\"a}fer}}, \bibinfo {author} {\bibfnamefont {W.~Z.}\ \bibnamefont
  {Van~Benschoten}}, \bibinfo {author} {\bibfnamefont {J.~J.}\ \bibnamefont
  {Shepherd}}, \ and\ \bibinfo {author} {\bibfnamefont {A.}~\bibnamefont
  {Gr{\"u}neis}},\ }\bibfield  {title} {\enquote {\bibinfo {title} {Sampling
  the reciprocal coulomb potential in finite anisotropic cells},}\ }\href@noop
  {} {\bibfield  {journal} {\bibinfo  {journal} {The Journal of Chemical
  Physics}\ }\textbf {\bibinfo {volume} {160}} (\bibinfo {year}
  {2024})}\BibitemShut {NoStop}%
\bibitem [{\citenamefont {Irmler}\ \emph {et~al.}(2018)\citenamefont {Irmler},
  \citenamefont {Burow},\ and\ \citenamefont {Pauly}}]{Irmler2018}%
  \BibitemOpen
  \bibfield  {author} {\bibinfo {author} {\bibfnamefont {A.}~\bibnamefont
  {Irmler}}, \bibinfo {author} {\bibfnamefont {A.~M.}\ \bibnamefont {Burow}}, \
  and\ \bibinfo {author} {\bibfnamefont {F.}~\bibnamefont {Pauly}},\ }\bibfield
   {title} {\enquote {\bibinfo {title} {Robust periodic fock exchange with
  atom-centered gaussian basis sets},}\ }\href {\doibase
  10.1021/acs.jctc.8b00122} {\bibfield  {journal} {\bibinfo  {journal} {Journal
  of Chemical Theory and Computation}\ }\textbf {\bibinfo {volume} {14}},\
  \bibinfo {pages} {4567} (\bibinfo {year} {2018})},\ \bibinfo {note} {pMID:
  30080979}\BibitemShut {NoStop}%
\bibitem [{\citenamefont {Kresse}\ and\ \citenamefont
  {Hafner}(1994)}]{kresse1994norm}%
  \BibitemOpen
  \bibfield  {author} {\bibinfo {author} {\bibfnamefont {G.}~\bibnamefont
  {Kresse}}\ and\ \bibinfo {author} {\bibfnamefont {J.}~\bibnamefont
  {Hafner}},\ }\bibfield  {title} {\enquote {\bibinfo {title} {Norm-conserving
  and ultrasoft pseudopotentials for first-row and transition elements},}\
  }\href@noop {} {\bibfield  {journal} {\bibinfo  {journal} {Journal of
  Physics: Condensed Matter}\ }\textbf {\bibinfo {volume} {6}},\ \bibinfo
  {pages} {8245} (\bibinfo {year} {1994})}\BibitemShut {NoStop}%
\bibitem [{\citenamefont {Kresse}\ and\ \citenamefont
  {Furthm{\"u}ller}(1996)}]{kresse1996efficient}%
  \BibitemOpen
  \bibfield  {author} {\bibinfo {author} {\bibfnamefont {G.}~\bibnamefont
  {Kresse}}\ and\ \bibinfo {author} {\bibfnamefont {J.}~\bibnamefont
  {Furthm{\"u}ller}},\ }\bibfield  {title} {\enquote {\bibinfo {title}
  {Efficient iterative schemes for ab initio total-energy calculations using a
  plane-wave basis set},}\ }\href@noop {} {\bibfield  {journal} {\bibinfo
  {journal} {Physical review B}\ }\textbf {\bibinfo {volume} {54}},\ \bibinfo
  {pages} {11169} (\bibinfo {year} {1996})}\BibitemShut {NoStop}%
\bibitem [{\citenamefont {Blum}\ \emph {et~al.}(2009)\citenamefont {Blum},
  \citenamefont {Gehrke}, \citenamefont {Hanke}, \citenamefont {Havu},
  \citenamefont {Havu}, \citenamefont {Ren}, \citenamefont {Reuter},\ and\
  \citenamefont {Scheffler}}]{blum2009ab}%
  \BibitemOpen
  \bibfield  {author} {\bibinfo {author} {\bibfnamefont {V.}~\bibnamefont
  {Blum}}, \bibinfo {author} {\bibfnamefont {R.}~\bibnamefont {Gehrke}},
  \bibinfo {author} {\bibfnamefont {F.}~\bibnamefont {Hanke}}, \bibinfo
  {author} {\bibfnamefont {P.}~\bibnamefont {Havu}}, \bibinfo {author}
  {\bibfnamefont {V.}~\bibnamefont {Havu}}, \bibinfo {author} {\bibfnamefont
  {X.}~\bibnamefont {Ren}}, \bibinfo {author} {\bibfnamefont {K.}~\bibnamefont
  {Reuter}}, \ and\ \bibinfo {author} {\bibfnamefont {M.}~\bibnamefont
  {Scheffler}},\ }\bibfield  {title} {\enquote {\bibinfo {title} {Ab initio
  molecular simulations with numeric atom-centered orbitals},}\ }\href@noop {}
  {\bibfield  {journal} {\bibinfo  {journal} {Computer Physics Communications}\
  }\textbf {\bibinfo {volume} {180}},\ \bibinfo {pages} {2175} (\bibinfo {year}
  {2009})}\BibitemShut {NoStop}%
\bibitem [{\citenamefont {Moerman}\ \emph {et~al.}(2022)\citenamefont
  {Moerman}, \citenamefont {Hummel}, \citenamefont {Gr{\"u}neis}, \citenamefont
  {Irmler},\ and\ \citenamefont {Scheffler}}]{moerman2022interface}%
  \BibitemOpen
  \bibfield  {author} {\bibinfo {author} {\bibfnamefont {E.}~\bibnamefont
  {Moerman}}, \bibinfo {author} {\bibfnamefont {F.}~\bibnamefont {Hummel}},
  \bibinfo {author} {\bibfnamefont {A.}~\bibnamefont {Gr{\"u}neis}}, \bibinfo
  {author} {\bibfnamefont {A.}~\bibnamefont {Irmler}}, \ and\ \bibinfo {author}
  {\bibfnamefont {M.}~\bibnamefont {Scheffler}},\ }\bibfield  {title} {\enquote
  {\bibinfo {title} {Interface to high-performance periodic coupled-cluster
  theory calculations with atom-centered, localized basis functions},}\
  }\href@noop {} {\bibfield  {journal} {\bibinfo  {journal} {The Journal of
  Open Source Software}\ }\textbf {\bibinfo {volume} {7}} (\bibinfo {year}
  {2022})}\BibitemShut {NoStop}%
\bibitem [{\citenamefont {Becke}(1992)}]{becke1992density}%
  \BibitemOpen
  \bibfield  {author} {\bibinfo {author} {\bibfnamefont {A.~D.}\ \bibnamefont
  {Becke}},\ }\bibfield  {title} {\enquote {\bibinfo {title}
  {Density-functional thermochemistry. i. the effect of the exchange-only
  gradient correction},}\ }\href@noop {} {\bibfield  {journal} {\bibinfo
  {journal} {The Journal of chemical physics}\ }\textbf {\bibinfo {volume}
  {96}},\ \bibinfo {pages} {2155} (\bibinfo {year} {1992})}\BibitemShut
  {NoStop}%
\bibitem [{\citenamefont {Lee}\ \emph {et~al.}(1988)\citenamefont {Lee},
  \citenamefont {Yang},\ and\ \citenamefont {Parr}}]{lee1988development}%
  \BibitemOpen
  \bibfield  {author} {\bibinfo {author} {\bibfnamefont {C.}~\bibnamefont
  {Lee}}, \bibinfo {author} {\bibfnamefont {W.}~\bibnamefont {Yang}}, \ and\
  \bibinfo {author} {\bibfnamefont {R.~G.}\ \bibnamefont {Parr}},\ }\bibfield
  {title} {\enquote {\bibinfo {title} {Development of the colle-salvetti
  correlation-energy formula into a functional of the electron density},}\
  }\href@noop {} {\bibfield  {journal} {\bibinfo  {journal} {Physical review
  B}\ }\textbf {\bibinfo {volume} {37}},\ \bibinfo {pages} {785} (\bibinfo
  {year} {1988})}\BibitemShut {NoStop}%
\bibitem [{\citenamefont {Hirata}\ \emph {et~al.}(1998)\citenamefont {Hirata},
  \citenamefont {Torii},\ and\ \citenamefont {Tasumi}}]{hirata1998density}%
  \BibitemOpen
  \bibfield  {author} {\bibinfo {author} {\bibfnamefont {S.}~\bibnamefont
  {Hirata}}, \bibinfo {author} {\bibfnamefont {H.}~\bibnamefont {Torii}}, \
  and\ \bibinfo {author} {\bibfnamefont {M.}~\bibnamefont {Tasumi}},\
  }\bibfield  {title} {\enquote {\bibinfo {title} {Density-functional crystal
  orbital study on the structures and energetics of polyacetylene isomers},}\
  }\href@noop {} {\bibfield  {journal} {\bibinfo  {journal} {Physical Review
  B}\ }\textbf {\bibinfo {volume} {57}},\ \bibinfo {pages} {11994} (\bibinfo
  {year} {1998})}\BibitemShut {NoStop}%
\bibitem [{\citenamefont {Yannoni}\ and\ \citenamefont
  {Clarke}(1983)}]{yannoni1983molecular}%
  \BibitemOpen
  \bibfield  {author} {\bibinfo {author} {\bibfnamefont {C.}~\bibnamefont
  {Yannoni}}\ and\ \bibinfo {author} {\bibfnamefont {T.}~\bibnamefont
  {Clarke}},\ }\bibfield  {title} {\enquote {\bibinfo {title} {Molecular
  geometry of cis-and trans-polyacetylene by nutation nmr spectroscopy},}\
  }\href@noop {} {\bibfield  {journal} {\bibinfo  {journal} {Physical review
  letters}\ }\textbf {\bibinfo {volume} {51}},\ \bibinfo {pages} {1191}
  (\bibinfo {year} {1983})}\BibitemShut {NoStop}%
\bibitem [{\citenamefont {Zhang}\ \emph {et~al.}(2013)\citenamefont {Zhang},
  \citenamefont {Ren}, \citenamefont {Rinke}, \citenamefont {Blum},\ and\
  \citenamefont {Scheffler}}]{zhang2013numeric}%
  \BibitemOpen
  \bibfield  {author} {\bibinfo {author} {\bibfnamefont {I.~Y.}\ \bibnamefont
  {Zhang}}, \bibinfo {author} {\bibfnamefont {X.}~\bibnamefont {Ren}}, \bibinfo
  {author} {\bibfnamefont {P.}~\bibnamefont {Rinke}}, \bibinfo {author}
  {\bibfnamefont {V.}~\bibnamefont {Blum}}, \ and\ \bibinfo {author}
  {\bibfnamefont {M.}~\bibnamefont {Scheffler}},\ }\bibfield  {title} {\enquote
  {\bibinfo {title} {Numeric atom-centered-orbital basis sets with
  valence-correlation consistency from h to ar},}\ }\href@noop {} {\bibfield
  {journal} {\bibinfo  {journal} {New Journal of Physics}\ }\textbf {\bibinfo
  {volume} {15}},\ \bibinfo {pages} {123033} (\bibinfo {year}
  {2013})}\BibitemShut {NoStop}%
\bibitem [{\citenamefont {Zhang}\ \emph {et~al.}(2019)\citenamefont {Zhang},
  \citenamefont {Logsdail}, \citenamefont {Ren}, \citenamefont {Levchenko},
  \citenamefont {Ghiringhelli},\ and\ \citenamefont
  {Scheffler}}]{zhang2019main}%
  \BibitemOpen
  \bibfield  {author} {\bibinfo {author} {\bibfnamefont {I.~Y.}\ \bibnamefont
  {Zhang}}, \bibinfo {author} {\bibfnamefont {A.~J.}\ \bibnamefont {Logsdail}},
  \bibinfo {author} {\bibfnamefont {X.}~\bibnamefont {Ren}}, \bibinfo {author}
  {\bibfnamefont {S.~V.}\ \bibnamefont {Levchenko}}, \bibinfo {author}
  {\bibfnamefont {L.}~\bibnamefont {Ghiringhelli}}, \ and\ \bibinfo {author}
  {\bibfnamefont {M.}~\bibnamefont {Scheffler}},\ }\bibfield  {title} {\enquote
  {\bibinfo {title} {Main-group test set for materials science and engineering
  with user-friendly graphical tools for error analysis: systematic benchmark
  of the numerical and intrinsic errors in state-of-the-art
  electronic-structure approximations},}\ }\href@noop {} {\bibfield  {journal}
  {\bibinfo  {journal} {New Journal of Physics}\ }\textbf {\bibinfo {volume}
  {21}},\ \bibinfo {pages} {013025} (\bibinfo {year} {2019})}\BibitemShut
  {NoStop}%
\bibitem [{\citenamefont {Levchenko}\ \emph {et~al.}(2015)\citenamefont
  {Levchenko}, \citenamefont {Ren}, \citenamefont {Wieferink}, \citenamefont
  {Johanni}, \citenamefont {Rinke}, \citenamefont {Blum},\ and\ \citenamefont
  {Scheffler}}]{levchenko2015hybrid}%
  \BibitemOpen
  \bibfield  {author} {\bibinfo {author} {\bibfnamefont {S.~V.}\ \bibnamefont
  {Levchenko}}, \bibinfo {author} {\bibfnamefont {X.}~\bibnamefont {Ren}},
  \bibinfo {author} {\bibfnamefont {J.}~\bibnamefont {Wieferink}}, \bibinfo
  {author} {\bibfnamefont {R.}~\bibnamefont {Johanni}}, \bibinfo {author}
  {\bibfnamefont {P.}~\bibnamefont {Rinke}}, \bibinfo {author} {\bibfnamefont
  {V.}~\bibnamefont {Blum}}, \ and\ \bibinfo {author} {\bibfnamefont
  {M.}~\bibnamefont {Scheffler}},\ }\bibfield  {title} {\enquote {\bibinfo
  {title} {Hybrid functionals for large periodic systems in an all-electron,
  numeric atom-centered basis framework},}\ }\href@noop {} {\bibfield
  {journal} {\bibinfo  {journal} {Computer Physics Communications}\ }\textbf
  {\bibinfo {volume} {192}},\ \bibinfo {pages} {60} (\bibinfo {year}
  {2015})}\BibitemShut {NoStop}%
\bibitem [{\citenamefont {Windom}\ \emph {et~al.}(2022)\citenamefont {Windom},
  \citenamefont {Perera},\ and\ \citenamefont
  {Bartlett}}]{windom2022examining}%
  \BibitemOpen
  \bibfield  {author} {\bibinfo {author} {\bibfnamefont {Z.~W.}\ \bibnamefont
  {Windom}}, \bibinfo {author} {\bibfnamefont {A.}~\bibnamefont {Perera}}, \
  and\ \bibinfo {author} {\bibfnamefont {R.~J.}\ \bibnamefont {Bartlett}},\
  }\bibfield  {title} {\enquote {\bibinfo {title} {Examining fundamental and
  excitation gaps at the thermodynamic limit: A combined (qtp) dft and coupled
  cluster study on trans-polyacetylene and polyacene},}\ }\href@noop {}
  {\bibfield  {journal} {\bibinfo  {journal} {The Journal of Chemical Physics}\
  }\textbf {\bibinfo {volume} {156}} (\bibinfo {year} {2022})}\BibitemShut
  {NoStop}%
\bibitem [{\citenamefont {Masios}\ \emph {et~al.}(2023)\citenamefont {Masios},
  \citenamefont {Irmler}, \citenamefont {Sch\"afer},\ and\ \citenamefont
  {Gr\"uneis}}]{masios2023}%
  \BibitemOpen
  \bibfield  {author} {\bibinfo {author} {\bibfnamefont {N.}~\bibnamefont
  {Masios}}, \bibinfo {author} {\bibfnamefont {A.}~\bibnamefont {Irmler}},
  \bibinfo {author} {\bibfnamefont {T.}~\bibnamefont {Sch\"afer}}, \ and\
  \bibinfo {author} {\bibfnamefont {A.}~\bibnamefont {Gr\"uneis}},\ }\bibfield
  {title} {\enquote {\bibinfo {title} {Averting the infrared catastrophe in the
  gold standard of quantum chemistry},}\ }\href {\doibase
  10.1103/PhysRevLett.131.186401} {\bibfield  {journal} {\bibinfo  {journal}
  {Phys. Rev. Lett.}\ }\textbf {\bibinfo {volume} {131}},\ \bibinfo {pages}
  {186401} (\bibinfo {year} {2023})}\BibitemShut {NoStop}%
\bibitem [{\citenamefont {Mattuck}(1992)}]{mattuck}%
  \BibitemOpen
  \bibfield  {author} {\bibinfo {author} {\bibfnamefont {R.~D.}\ \bibnamefont
  {Mattuck}},\ }\href@noop {} {\emph {\bibinfo {title} {A guide to {Feynman}
  diagrams in the many-body problem}}},\ \bibinfo {edition} {2nd}\ ed.,\ Dover
  books on physics and chemistry\ (\bibinfo  {publisher} {Dover Publications},\
  \bibinfo {address} {New York},\ \bibinfo {year} {1992})\BibitemShut {NoStop}%
\bibitem [{\citenamefont {Bintrim}\ and\ \citenamefont
  {Berkelbach}(2021)}]{bintrim2021full}%
  \BibitemOpen
  \bibfield  {author} {\bibinfo {author} {\bibfnamefont {S.~J.}\ \bibnamefont
  {Bintrim}}\ and\ \bibinfo {author} {\bibfnamefont {T.~C.}\ \bibnamefont
  {Berkelbach}},\ }\bibfield  {title} {\enquote {\bibinfo {title}
  {Full-frequency gw without frequency},}\ }\href@noop {} {\bibfield  {journal}
  {\bibinfo  {journal} {The Journal of Chemical Physics}\ }\textbf {\bibinfo
  {volume} {154}} (\bibinfo {year} {2021})}\BibitemShut {NoStop}%
\bibitem [{\citenamefont {Nooijen}\ and\ \citenamefont
  {Snijders}(1992)}]{nooijen1992coupled}%
  \BibitemOpen
  \bibfield  {author} {\bibinfo {author} {\bibfnamefont {M.}~\bibnamefont
  {Nooijen}}\ and\ \bibinfo {author} {\bibfnamefont {J.~G.}\ \bibnamefont
  {Snijders}},\ }\bibfield  {title} {\enquote {\bibinfo {title} {Coupled
  cluster approach to the single-particle green's function},}\ }\href@noop {}
  {\bibfield  {journal} {\bibinfo  {journal} {International Journal of Quantum
  Chemistry}\ }\textbf {\bibinfo {volume} {44}},\ \bibinfo {pages} {55}
  (\bibinfo {year} {1992})}\BibitemShut {NoStop}%
\bibitem [{\citenamefont {Backhouse}\ and\ \citenamefont
  {Booth}(2022)}]{backhouse2022constructing}%
  \BibitemOpen
  \bibfield  {author} {\bibinfo {author} {\bibfnamefont {O.~J.}\ \bibnamefont
  {Backhouse}}\ and\ \bibinfo {author} {\bibfnamefont {G.~H.}\ \bibnamefont
  {Booth}},\ }\bibfield  {title} {\enquote {\bibinfo {title} {Constructing
  “full-frequency” spectra via moment constraints for coupled cluster
  green’s functions},}\ }\href@noop {} {\bibfield  {journal} {\bibinfo
  {journal} {Journal of Chemical Theory and Computation}\ }\textbf {\bibinfo
  {volume} {18}},\ \bibinfo {pages} {6622} (\bibinfo {year}
  {2022})}\BibitemShut {NoStop}%
\bibitem [{\citenamefont {Vosko}\ \emph {et~al.}(1980)\citenamefont {Vosko},
  \citenamefont {Wilk},\ and\ \citenamefont {Nusair}}]{vosko1980accurate}%
  \BibitemOpen
  \bibfield  {author} {\bibinfo {author} {\bibfnamefont {S.~H.}\ \bibnamefont
  {Vosko}}, \bibinfo {author} {\bibfnamefont {L.}~\bibnamefont {Wilk}}, \ and\
  \bibinfo {author} {\bibfnamefont {M.}~\bibnamefont {Nusair}},\ }\bibfield
  {title} {\enquote {\bibinfo {title} {Accurate spin-dependent electron liquid
  correlation energies for local spin density calculations: a critical
  analysis},}\ }\href@noop {} {\bibfield  {journal} {\bibinfo  {journal}
  {Canadian Journal of physics}\ }\textbf {\bibinfo {volume} {58}},\ \bibinfo
  {pages} {1200} (\bibinfo {year} {1980})}\BibitemShut {NoStop}%
\bibitem [{\citenamefont {Massidda}\ \emph {et~al.}(1993)\citenamefont
  {Massidda}, \citenamefont {Posternak},\ and\ \citenamefont
  {Baldereschi}}]{massidda1993hartree}%
  \BibitemOpen
  \bibfield  {author} {\bibinfo {author} {\bibfnamefont {S.}~\bibnamefont
  {Massidda}}, \bibinfo {author} {\bibfnamefont {M.}~\bibnamefont {Posternak}},
  \ and\ \bibinfo {author} {\bibfnamefont {A.}~\bibnamefont {Baldereschi}},\
  }\bibfield  {title} {\enquote {\bibinfo {title} {Hartree-fock lapw approach
  to the electronic properties of periodic systems},}\ }\href@noop {}
  {\bibfield  {journal} {\bibinfo  {journal} {Physical Review B}\ }\textbf
  {\bibinfo {volume} {48}},\ \bibinfo {pages} {5058} (\bibinfo {year}
  {1993})}\BibitemShut {NoStop}%
\bibitem [{\citenamefont {Quintero-Monsebaiz}\ \emph
  {et~al.}(2022)\citenamefont {Quintero-Monsebaiz}, \citenamefont {Monino},
  \citenamefont {Marie},\ and\ \citenamefont {Loos}}]{quintero2022connections}%
  \BibitemOpen
  \bibfield  {author} {\bibinfo {author} {\bibfnamefont {R.}~\bibnamefont
  {Quintero-Monsebaiz}}, \bibinfo {author} {\bibfnamefont {E.}~\bibnamefont
  {Monino}}, \bibinfo {author} {\bibfnamefont {A.}~\bibnamefont {Marie}}, \
  and\ \bibinfo {author} {\bibfnamefont {P.-F.}\ \bibnamefont {Loos}},\
  }\bibfield  {title} {\enquote {\bibinfo {title} {Connections between
  many-body perturbation and coupled-cluster theories},}\ }\href@noop {}
  {\bibfield  {journal} {\bibinfo  {journal} {The Journal of Chemical Physics}\
  }\textbf {\bibinfo {volume} {157}} (\bibinfo {year} {2022})}\BibitemShut
  {NoStop}%
\bibitem [{\citenamefont {Scuseria}\ \emph {et~al.}(2008)\citenamefont
  {Scuseria}, \citenamefont {Henderson},\ and\ \citenamefont
  {Sorensen}}]{scuseria2008ground}%
  \BibitemOpen
  \bibfield  {author} {\bibinfo {author} {\bibfnamefont {G.~E.}\ \bibnamefont
  {Scuseria}}, \bibinfo {author} {\bibfnamefont {T.~M.}\ \bibnamefont
  {Henderson}}, \ and\ \bibinfo {author} {\bibfnamefont {D.~C.}\ \bibnamefont
  {Sorensen}},\ }\bibfield  {title} {\enquote {\bibinfo {title} {The ground
  state correlation energy of the random phase approximation from a ring
  coupled cluster doubles approach},}\ }\href@noop {} {\bibfield  {journal}
  {\bibinfo  {journal} {The Journal of chemical physics}\ }\textbf {\bibinfo
  {volume} {129}} (\bibinfo {year} {2008})}\BibitemShut {NoStop}%
\bibitem [{\citenamefont {Yan}\ \emph {et~al.}(2000)\citenamefont {Yan},
  \citenamefont {Perdew},\ and\ \citenamefont {Kurth}}]{yan2000density}%
  \BibitemOpen
  \bibfield  {author} {\bibinfo {author} {\bibfnamefont {Z.}~\bibnamefont
  {Yan}}, \bibinfo {author} {\bibfnamefont {J.~P.}\ \bibnamefont {Perdew}}, \
  and\ \bibinfo {author} {\bibfnamefont {S.}~\bibnamefont {Kurth}},\ }\bibfield
   {title} {\enquote {\bibinfo {title} {Density functional for short-range
  correlation: Accuracy of the random-phase approximation for isoelectronic
  energy changes},}\ }\href@noop {} {\bibfield  {journal} {\bibinfo  {journal}
  {Physical Review B}\ }\textbf {\bibinfo {volume} {61}},\ \bibinfo {pages}
  {16430} (\bibinfo {year} {2000})}\BibitemShut {NoStop}%
\bibitem [{\citenamefont {Sch\"afer}\ \emph {et~al.}(2024)\citenamefont
  {Sch\"afer}, \citenamefont {Van~Benschoten}, \citenamefont {Shepherd},\ and\
  \citenamefont {Gr\"uneis}}]{Schaefer2024}%
  \BibitemOpen
  \bibfield  {author} {\bibinfo {author} {\bibfnamefont {T.}~\bibnamefont
  {Sch\"afer}}, \bibinfo {author} {\bibfnamefont {W.~Z.}\ \bibnamefont
  {Van~Benschoten}}, \bibinfo {author} {\bibfnamefont {J.~J.}\ \bibnamefont
  {Shepherd}}, \ and\ \bibinfo {author} {\bibfnamefont {A.}~\bibnamefont
  {Gr\"uneis}},\ }\bibfield  {title} {\enquote {\bibinfo {title} {{Sampling the
  reciprocal Coulomb potential in finite anisotropic cells}},}\ }\href
  {\doibase 10.1063/5.0182729} {\bibfield  {journal} {\bibinfo  {journal} {The
  Journal of Chemical Physics}\ }\textbf {\bibinfo {volume} {160}},\ \bibinfo
  {pages} {051101} (\bibinfo {year} {2024})}\BibitemShut {NoStop}%
\end{thebibliography}%

\end{document}